\documentclass[twocolumn]{article}
\usepackage{amssymb}
\usepackage{cite}
\usepackage{graphicx}
\usepackage[T2A]{fontenc}
\usepackage[cp866nav]{inputenc}
\newcommand{\Sp}{\mathop{\rm Sp}\nolimits}
\newcommand{\D}{\mathrm{d}}
\newcommand{\E}{\mathrm{e}}
\newcommand{\I}{\mathrm{i}}
\renewcommand{\vec}[1]{\mathbf{#1}}

\begin{document}

\twocolumn[
\title{%{\large\null\hfill УДК: 537.61}\\ [1em]
Strong coupling approach in dynamical mean-field theory
for strongly correlated electron systems}

\author{Ihor V. Stasyuk and Andrij M. Shvaika\\ [1ex]
    Institute for Condensed Matter Physics\\
    of the National Academy of Sciences of Ukraine,\\
    1 Svientsitskii Str., UA--79011 Lviv, Ukraine}

\date{}

\maketitle ]

\begin{abstract}
We review two analytical approaches in Dynamical Mean-Field Theory (DMFT)
based on a perturbation theory expansion over the electron hopping to and
from the self consistent environment. In the first approach the effective
single impurity Anderson model (SIAM) is formulated in terms of the
auxiliary Fermi-fields and the projection (irreducible Green's function)
technique is used for its solution. A system of the DMFT equations is
obtained that includes as simple specific cases a number of known
approximations (Hubbard-III, AA, MAA, \dots). The second approach is based
on the diagrammatic technique (Wick's theorem) for Hubbard operators that
allows to construct a thermodynamically consistent theory when SIAM
exactly splits into four components (subspaces): two Fermi liquid and two
non-Fermi liquid. The results for the density of states, concentration
dependences of the band energies, chemical potential and magnetic order
parameters are presented for different self-consistent approximations (AA,
strong coupling Hartree--Fock and further).
\end{abstract}

\section{Introduction}

Many unconventional properties (e.g., metal--insulator transition,
electronic (an\-ti)fer\-ro\-mag\-ne\-tism) of the nar\-row-band systems
(transition metals and their compounds, some organic systems, high-$T_c$
superconductors, etc.) can be explained only by the proper treatment of
the strong local electron correlations. The simplest models allowing for
the electron correlations are a single-band Hubbard model with on-site
repulsion $U$ and hopping energy $t$ and its strong-coupling limit ($U\gg
t$): $t-J$ model. Recent studies of the Hubbard-type models connected
mainly with the theory of high-$T_c$ superconductivity and performed in
the weak- $(U\le 4t)$ and strong- $(U\gg t)$ coupling limits, elucidate
some important features of these models \cite{DagottoReview}. But still a
lot of problems remains, especially for the $U\gg t$ case where there are
no rigorous approaches.

Despite the relative simplicity of the models used for their description
the theory of electron spectrum and thermodynamic properties of such
systems is far from its final completion. The use of localized (atomic)
basis of electron states is the general feature of the models.
Corresponding Hamiltonians
\begin{equation}
  H=\sum_{i} H_{i}+\sum_{\langle ij\rangle} \sum_{\sigma}
  t_{ij}a_{i\sigma}^{\dagger} a_{j\sigma}
  \label{eq1.1}
\end{equation}
include, on the one hand the electron transfer (hopping) $t_{ij}$ between
neighbouring sites (atoms) in the crystal lattice and on the other hand
the short-range single-site electron correlations. It is primarily the
on-site energy of Coulomb repulsion $U$ in the case of the Hubbard model
and the models based on that one:
\begin{equation}
  H_{i} = U n_{i\uparrow}n_{i\downarrow} - \mu \sum_{\sigma} n_{i\sigma} \; .
  \label{eq1.2}
\end{equation}

Models like (\ref{eq1.1}), (\ref{eq1.2}) can be solved exactly in two
limiting cases: atomic limit ($t=0$) and band electrons ($U=0$). Near
these extreme cases the expansions in terms of $t$ or $U$ are used, but
the consistent formulation of the perturbation theory especially in the
case of strong coupling is not a simple task. The case of an intermediate
coupling $t\sim U$ is more complicated for consideration. In this region
of parameter values, the splitting in the band electron spectrum and the
metal--insulator transition takes place.

Due to the presence of strong electron correlations, the state of the
electron system and its properties depend essentially on the mean electron
concentration. At a different filling of electron states and depending on
the relation between $t$ and $U$ parameters, the system can be
paramagnetic or the transition into ferro- (antiferro-) magnetic phase can
take place. In the case of the more complicated structure of the
Hamiltonian $H_{i}$ (due to the allowance for the other, besides electron,
degrees of freedom) or when the interaction is extended to the nearest
neighbours in a lattice, the charge ordering can appear; the effects of
phase separation become  possible as well. The listed phenomena are the
subject of study in the framework of various approaches and methods.

A new impulse in the investigations in this field is connected with the
development of a new approach having its origin in works
\cite{MetznerVollhardt,MuellerHartmann,Metzner} where the study of the
(\ref{eq1.1}), (\ref{eq1.2})-type model in the limit of infinite
dimensionality of space ($d=\infty$) has been proposed. Due to the
principal simplifications in the perturbation series taking place in this
case the possibility exists to obtain  exact results using the scheme that
corresponds to the well known coherent potential approximation (CPA) in
the theory of disordered crystalline alloys. The  rapidly developing
corresponding method became known as the dynamical mean-field theory
(DMFT).

The central point in this method is formulation and solution of the
auxiliary single-site problem. An initial model is mapped on that one
while considering one site characteristics of the electron spectrum, such
as single-site electron Green's function (see
\cite{GeorgesKotliar,Jarrel,JanisVollhardt}, as well as the reviews
\cite{Izyumovreview,DMFTreview}). In this case the separated lattice site
is considered as placed in  some effective environment. Since the
processes of electron hopping from the atom and returning into the atom
are taken into account, the mean field acting on the electron states of
the atom possesses a dynamical nature. This field is described by the
coherent potential $J_{\sigma}(\omega)$ that should be determined in a
self-consistent way. The analytical properties of the solutions in the
DMFT are considered in \cite{Pruschke2001}.

Only in  some simple cases the single-site problem can be solved
analytically \cite{BrandtMielsch,ShvaikaSusc} (e.g., the Falicov--Kimball
(FK) model \cite{Falicov}). In general, including the Hubbard model, the
application of numerical or seminumerical (such as quantum Monte Carlo
\cite{Jarrel,Rozenberg1992,Georges1992} or exact diagonalization
\cite{Caffarel,Si1994} as well as numerical renormalization group
\cite{Bullareview}, see \cite{DMFTreview}) methods turns out to be
necessary.

At present, the Dynamical Mean-Field Theory is applied to investigate
different effects in the various systems described by the simplified or
realistic models. First applications were devoted to the investigation of
the changes of the density of states at the metal-insulator transition and
appearance of the antiferromagnetic and ferromagnetic states in the
Hubbard model
\cite{Jarrel,Georges1992,Georges1993,Pruschke1993,Pruschke,Kopietz,T-U2,%
Vollhardt1997,Herrmann1997,Noack1999,Schlipf1999,Bulla1999,Bulla1999a} and
other strongly correlated electron models: Hubbard model with orbiral
degeneracy \cite{Momoi1998} and disorder \cite{Denteneer}, boson-fermion
model \cite{Robin}, extended Hubbard model \cite{Pietig}, double-exchange
model \cite{Nagai}, Falicov--Kimball model with correlated hopping
\cite{Schiller}, two-band Hubbard model \cite{Ono,Ohashi}, periodic
Anderson model \cite{Held2000}.

Besides, different types of the response functions and transport
coefficients are also calculated. The general basics how to derive
response functions in DMFT are given in
\cite{BrandtMielsch,Zlatic,ShvaikaSusc}. Investigations of the charge and
magnetic susceptibilities revealed also chess-board charge-density-wave
phase at half-filling \cite{BrandtMielsch} as well as incommensurate order
and phase separation at other fillings
\cite{Freericks1993,Freericks1993a}. DMFT is used also to investigate
optical conductivity \cite{Pruschke1993,Pruschke,Scheweitzer,Moeller},
electronic Raman scattering \cite{Freericks2001,Freericks2001a},
thermoelectric response \cite{Palsson}.

In last years, DMFT is used as an approximation scheme to consider the
electron-electron interaction together with band degeneracy and lattice
structure of the actual materials within the so-called LDA+DMFT approach,
that allows to describe correctly the insulating state of the
transition-metal oxides, band structure and phase diagrams of the
different compounds \cite{Held2001,Held2001a}.

At the same time it is of interest to develop approximate analytic
approaches to the solution of the single-site problem. Their application
at that stage is more effective than at considering the full model (a
short review of  such attempts was given recently in
\cite{PotthoffHWN,Gebhard}). The availability of the analytical (even of
approximate) method is useful especially for  new models as well as at the
transition to the finite dimensionality of the system. The accuracy of
approximation can be estimated relating to the results of numerical
calculations.

The first analytical approximation proposed for the Hubbard model was a
simple Hubbard-I approximation \cite{H-I} (see Ref.~\cite{Dorneich} for
its possible improvement) which is correct in the atomic ($t=0$) and band
($U=0$) limits but is inconsistent in the intermediate cases and cannot
describe the metal--insulator transition. Hubbard's alloy-analogy solution
\cite{H-III} (so-called Hubbard-III approximation) incorporates into the
theory an electron scattering on the charge and spin fluctuations that
allows us to give qualitative description of the changes of the
density-of-state at the metal--insulator transition point. Hubbard-I and
Hubbard-III approximations introduce two types of particles (electrons
moving between empty sites and electrons moving between sites occupied by
electrons of opposite spin) with the different energies that differ by $U$
and form two Hubbard bands. Related schemes of the so-called two-pole
approximations \cite{Roth,Nolting}, which are justified by the $t/U \ll 1$
perturbation theory expansions \cite{H_L}, are also considered. However,
in the recent QMC studies \cite{Grober,Pairault} there are clearly
distinguished four bands in the spectral functions rather than the two
bands predicted by the two-pole approximations. Such four-band structure
is reproduced by the strong-coupling expansion for the Hubbard model
\cite{Pairault} in the one-dimensional case. There are also analytical
approximations developed specially for the effective environment in the
DMFT \cite{Bulla2000,Potthoff2001}. Within other approaches let us mention
non-crossing approximation \cite{Pruschke,Obermeier}, Edwards--Hertz
approach \cite{Edwards,Wermbter}, iterative perturbation theory
\cite{Kajueter,Wegner}, alloy-analogy based approaches
\cite{Herrmann,Potthoff}, and linked cluster expansions
\cite{Metzner,Janis}, which are reliable in certain limits and the
construction of the thermodynamically consistent theory still remains open
\cite{Gebhard}.

The aim of this paper is to review two recently proposed approaches
\cite{StasyukCMP,ShvaikaPRB} based on the rigorous perturbation theory
scheme in terms of electron hopping for the Hubbard-type models.

The first approach \cite{StasyukCMP} is based on the technique of the
irreducible Green's functions. The procedure of projecting onto the basic
set of operators is used (the set consists of the single-site electron
Hubbard operators of the Fermi-type). The recipe is given for the
construction of the system of equations for the coherent potential and
self-consistency parameter (having the meaning of a static part of the
effective internal field) in the approach that is a generalization of the
Hubbard-III approximation. Specific cases are considered corresponding to
the more simple approximations of the alloy-analogy (AA) or modified
alloy-analogy (MAA) type \cite{Gebhard,Herrmann} in the DMFT method as
well as to the certain decoupling procedure in the two-time Green's
function method when applied to the initial electron problem.

Another possibility is to build analytical approaches by the systematic
perturbation expansion in terms of the electron hopping
\cite{Cojocaru,Moskalenko,IzyumovCMP} using diagrammatic technique for
Hubbard operators \cite{Slobodyan,IzyumovBook}. One of them was proposed
for the Hubbard ($U=\infty$ limit) and $t-J$ models
\cite{IzyumovLetfulov,IzyumovPRB}. The lack of such approach is connected
with the concept of a ``hierarchy'' system for Hubbard operators when the
form of the diagrammatic series and final results strongly depend on the
system of the pairing priority for Hubbard operators. On the other hand it
is difficult to generalize it on the case of the arbitrary $U$.

In the second part of this paper we show how a rigorous perturbation
theory scheme in terms of electron hopping that is based on the Wick's
theorem for Hubbard operators \cite{Slobodyan,IzyumovBook} and is valid
for arbitrary value of $U$ ($U<\infty$) and does not depend on the
``hierarchy'' system for $X$ operators can be developed for the
Hubbard-type models \cite{ShvaikaPRB}. In the limit of infinite spatial
dimensions, these analytical schemes allow us to build a self-consistent
Baym--Kadanoff type theory \cite{BaymKadanoff,Baym} for the Hubbard model
and some analytical results are given for simple approximations. The
Falicov--Kimball model is also considered as an exactly solvable limit of
Hubbard model.

\section{Hubbard model and similar models in a limit of infinite dimension
of space ($d=\infty$)}

The transition to the $d=\infty$ limit in the DMFT approach is accompanied
by the scaling of the electron transfer parameter
\begin{equation}
  t=\frac{t^{*}}{\sqrt{d}}\;.
  \label{eq2.1}
\end{equation}
In the case of $d$-dimensional hypercubic lattice with an electron
spectrum
\begin{equation}
  \varepsilon_{k}=\frac{2t^{*}}{\sqrt{d}} \sum_{\alpha=1}^{d}
  \cos k_{\alpha} a\;,
  \label{eq2.2}
\end{equation}
this procedure leads to the Gaussian density of electron states
\cite{MetznerVollhardt}
\begin{equation}
  \rho_{0}(\omega) = \frac{1}{2\sqrt{\pi}t^{*}}
  \exp\left(-\frac{\omega^{2}}{4t^{*2}}\right) \; .
  \label{eq2.3}
\end{equation}
The average kinetic energy remains constant in this case in the limit
$d=\infty$.

The scaling (\ref{eq2.1}) has a significant effect on the structure of
diagrammatic series for single-electron Green's functions of the model of
the (\ref{eq1.1}) and (\ref{eq1.2}) type. In particular, the irreducible
self-energy part of such a function becomes a purely local (a single-site)
quantity \cite{MetznerVollhardt,MuellerHartmann}:
\begin{equation}
  \Sigma_{ij,\sigma}(\omega) = \Sigma_{\sigma}(\omega)\delta_{ij}\;,
  \qquad d=\infty \;.
  \label{eq2.4}
\end{equation}
The Fourier-transform of $\Sigma_{ij,\sigma}(\omega)$ is hence
momentum-independent
\begin{equation}
  \Sigma_{\sigma}(\vec{k},\omega) = \Sigma_{\sigma}(\omega)\;.
  \label{eq2.5}
\end{equation}
This leads to tremendous simplifications in all many-body calculations for
the Hubbard model and related models and enables us to obtain the exact
numerical results for the main parameters of the electron spectrum, to
describe magnetic phase transitions and the metal--insulator
transformation etc. (see, for example, \cite{DMFTreview,Gebhard}).

The possibility of obtaining  exact solutions in the $d=\infty$ limit
opens the way to the development of a theory based on the expansion in
powers of $1/d$ (the results for $d=\infty$ can be considered as zero
approximation in this case). Such approaches have been  elaborated for the
last few years \cite{DongenPRB,Hettler}. On the other hand,  consideration
in the framework in the $d=\infty$ limit is not only of an academic
interest. It turns out that a set of the known approximating schemes or
methods is correct in the $d=\infty$ limit. Besides, the obtained physical
conclusions can be transferred in many cases to the system with finite
dimensions keeping their suitability even at $d=3$.

The formal scheme of calculating  the electron Green's functions and the
main thermodynamical quantities can be developed basing on the
diagrammatic expansions in powers of interaction parameters (such as
energy $U$ in the case of the Hubbard model) or matrix elements of the
electron transfer $t_{ij}$. The electron Green's function in
($\vec{k},\omega$) representation
\begin{equation}
  G_{k}^{\sigma} (\omega) = \sum_{i-j}
  \E^{\I \vec{k}(\vec{R}_{i}-\vec{R}_{j})}
  G_{ij,\sigma} (\omega)
  \label{eq2.6}
\end{equation}
can be expressed in the first or in the second of these cases as
\begin{equation}
  G_{\vec{k}}^{\sigma}(\omega) = \frac{1}{\omega +\mu - t_{\vec{k}} -
  \Sigma_{\sigma}(\omega)}
  \label{eq2.7}
\end{equation}
or
\begin{equation}
  G_{\vec{k}}^{\sigma}(\omega) = \frac{1}{\Xi_{\sigma}^{-1}(\omega)-t_{\vec{k}}}\;,
  \label{eq2.8}
\end{equation}
where $\Sigma_{\sigma}(\omega)$ or $\Xi_{\sigma}(\omega)$ are the
irreducible parts (in the diagrammatic representation) according to Dyson
or Larkin, respectively,
\begin{equation}
  \Xi_{\sigma}^{-1}(\omega)=\omega +\mu - \Sigma_{\sigma}(\omega)\;.
\end{equation}

To calculate the $\Sigma_{\sigma}(\omega)$ [or $\Xi_{\sigma}(\omega)$]
function, the effective single-site problem is used. As was shown in
\cite{BrandtMielsch}, the transition to this problem corresponds to the
replacement
\begin{eqnarray}
&&  \E^{-\beta H} \to \E^{-\beta H_{\mathrm{eff}}}
 =  \E^{-\beta
  H_{0}}
  \label{eq2.9}
  \\
&&  \times  T\exp \left\{
  \! {-}\int_{0}^{\beta} \!\!\!\D\tau \int_{0}^{\beta} \!\!\!\D\tau' \sum_{\sigma}
  J_{\sigma}(\tau{-}\tau') a_{\sigma}^{\dagger}(\tau) a_{\sigma}(\tau')\right\} ,
\nonumber
\end{eqnarray}
where
\begin{equation}
  H_{0}=H_{i}
  \label{eq2.10}
\end{equation}
and $J_{\sigma}(\tau-\tau')$ is an effective auxiliary field which is
determined self-consistently from the condition that the same irreducible
part $\Xi_{\sigma}(\omega)$ determines the lattice function (\ref{eq2.8})
as well as the Green's function $G_{\sigma}^{(a)}(\omega)$ of the
effective single-site problem. The last one is connected with
$\Xi_{\sigma}(\omega)$ and $J_{\sigma}(\omega)$ by the relation
\begin{equation}
  G_{\sigma}^{(a)}(\omega) = \frac{1}{\Xi_{\sigma}^{-1}(\omega) -
  J_{\sigma}(\omega)}\;.
  \label{eq2.11}
\end{equation}

On the other hand,
\begin{equation}
  G_{\sigma}^{(a)} (\omega) = G_{ii,\sigma}(\omega) = \frac{1}{N}
  \sum_{\vec{k}}
  G_{\vec{k}}^{\sigma} (\omega)\;.
  \label{eq2.12}
\end{equation}

Dynamical field $J_{\sigma}(\tau-\tau')$ describes electron hopping from
the given site into the environment and vice versa; the electron
propagates in the environment without going through this site between
moments $\tau$ and $\tau'$. The expression
\begin{equation}
  J_{\sigma}(\omega) = \sum_{kj} t_{ik}t_{ij} G_{kj,\sigma}^{(i)}
  \label{eq2.13}
\end{equation}
corresponds to this situation (the relation (\ref{eq2.13}) is known from
the standard CPA scheme \cite{Velicki,Ehrenreich}); here
$G_{kj,\sigma}^{(i)}$ is the electron Green's function for a crystal with
the removed site $i$.

The set of equations (\ref{eq2.8}), (\ref{eq2.11}) and (\ref{eq2.12})
becomes closed when it is supplemented by the functional dependence
\begin{equation}
  G_{\sigma}^{(a)} (\omega) = f([J_{\sigma}(\omega)])\;,
  \label{eq2.14}
\end{equation}
which is obtained as the result of  solving the effective single-site
problem with the statistical operator $\exp(-\beta H_{\mathrm{eff}})$. It
is possible to do this in an analytical way only in  some cases of simple
models (a Falicov--Kimball model \cite{BrandtMielsch}; a
pseudospin--electron model at $U=0$ \cite{ShvaikaJPS}; a usual binary
alloy model). In general, numerical methods are used.

The scheme described lies at the basis of the above mentioned DMFT
approach used in the last years in considering strongly correlated
electron systems.

\section{Electron Green's functions of the effective single-site
problem}

As it was mentioned, the central point in the DMFT approach is the
solution of the effective single-site problem and the determination of the
connection between the dynamical mean field (coherent potential)
$J_{\sigma}(\omega)$ and the single-site electron Green's function
$G_{\sigma}^{(a)}(\omega)$. Recently an approximate scheme
\cite{StasyukCMP}, which is based on the technique of the irreducible
two-time temperature Green's functions and leads to the results having an
interpolating character, was proposed. The Hubbard model is taken into
consideration to illustrate the method.

Let us reformulate a single-site problem introducing explicitly an
effective Hamiltonian
\begin{equation}
  \tilde{H}_{\mathrm{eff}} = H_{0}+ V\sum_{\sigma} (a_{\sigma}^{\dagger} \xi_{\sigma} +
  \xi_{\sigma}^{\dagger} a_{\sigma})+H_{\xi}\;,
  \label{eq3.1}
\end{equation}
where the auxiliary Fermi-field $(\xi_{\sigma},\xi_{\sigma}^{\dagger})$ is
brought in. It describes the environment of the selected site and formally
is characterized by the Hamiltonian $H_{\xi}$. The single-electron
transitions between the site and the environment are taken into account.

An explicit form of the Hamiltonian $H_{\xi}$ is unknown. Let us consider,
however, the Green's function
\begin{equation}
  {\cal{G}}_{\sigma}(\omega) = \langle\langle \xi_{\sigma}
  |\xi_{\sigma}^{\dagger}\rangle\rangle_{\omega}^{(H_{\xi})}
  \label{eq3.2}
\end{equation}
for auxiliary fermions as the given function. The Green's function
${\cal{G}}_{\sigma}(\tau{-}\tau')=\langle
T_{\tau}\xi_{\sigma}^{\dagger}(\tau)\xi_{\sigma}(\tau')\rangle_{\xi}$
[where averaging is performed with the part $H_{\xi}$ of the Hamiltonian
(\ref{eq3.1})] corresponds to the function (\ref{eq3.2}) in the
Matsubara's representation. It is shown in \cite{StasyukCMP} that the
expansion of the $\exp (-\beta\tilde{H}_{\mathrm{eff}})$ operator in
powers of $V$ and the subsequent averaging over the states of
$\xi$-subsystem using the Wick's theorem and functions (\ref{eq3.2}) leads
to the statistical operator (\ref{eq2.9}):
\begin{equation}
  \langle \exp(-\beta\tilde{H}_{\mathrm{eff}})\rangle^{(H_{\xi})} = \exp (-\beta
  H_{\mathrm{eff}})\;.
  \label{eq3.3}
\end{equation}
The relation
\begin{equation}
  2\pi V^{2}{\cal{G}}_{\sigma}(\omega) = J_{\sigma}(\omega)
  \label{eq3.4}
\end{equation}
takes place in this case.

The obtained result points out to the possibility of the Green's function
$G_{\sigma}^{(a)}(\omega)$ calculation based on the Hamiltonian
$\tilde{H}_{\mathrm{eff}}$. The averaging over the $a$,
$a^{\dagger}$-variables is performed with the use of the Gibbs
distribution while over the $\xi$, $\xi^{\dagger}$-variables it is done
with the help of function (\ref{eq3.2}).

Let us write the Hamiltonian (\ref{eq3.1}) for the case of the Hubbard
model in terms of Hubbard operators
\begin{eqnarray}
&&  \tilde{H}_{\mathrm{eff}} = -\mu \biggl(\sum\limits_{\sigma}
X^{\sigma\sigma} +
  2X^{22}\biggr) + UX^{22} +  \nonumber\\
&& \quad  + V\sum\limits_{\sigma} \left[(X^{\sigma 0} + \sigma
X^{2\bar\sigma})
  \xi_{\sigma} + \xi_{\sigma}^{\dagger} (X^{0\sigma} +\sigma X^{\bar\sigma
  2})\right]+
  \nonumber\\
&& \quad +   H_{\xi}\;.
  \label{eq3.5}
\end{eqnarray}
Here the basis of single-site states $|n_{\uparrow}n_{\downarrow}\rangle$
\begin{equation}
  \begin{array}{ll}
  |0\rangle = |0,0\rangle\;,\quad & |\downarrow\rangle = |0,1\rangle \;, \\
  |2\rangle = |1,1\rangle\;,\quad & |\uparrow\rangle = |1,0\rangle
  \end{array}
  \label{eq3.6}
\end{equation}
is used ($\sigma=\uparrow,\downarrow$). In this case the Green's function
$G_{\sigma}^{(a)}(\omega)$ can be written in the form
\begin{eqnarray}
  G_{\sigma}^{(a)} &=& \langle\langle X^{0\sigma} |X^{\sigma
  0}\rangle\rangle_{\omega} {+} \sigma\langle\langle X^{0\sigma}
  |X^{2\bar\sigma} \rangle\rangle_{\omega}\nonumber\\
  &+& \sigma \langle\langle
  X^{\bar\sigma 2} | X^{\sigma 0} \rangle\rangle_{\omega} {+} \langle\langle
  X^{\bar\sigma 2} |X^{2\bar\sigma}\rangle\rangle_{\omega}
  \label{eq3.7}
\end{eqnarray}
(a representation in terms of the two-time Green's functions is used).

We will write the equations for functions (\ref{eq3.7}) using the
equations of motion for $X$-operators:
\arraycolsep=1pt
\begin{eqnarray}
\arraycolsep=1pt
  \I \frac{\D}{\D t} X^{0\sigma} (t) &=& [X^{0\sigma}, \tilde{H}_{\mathrm{eff}}]
  = -\mu X^{0\sigma} + V(X^{00}+X^{\sigma\sigma})\xi_{\sigma}
   \nonumber\\
  &+& VX^{\bar\sigma \sigma} \xi_{\bar\sigma} +
  \sigma VX^{02} \xi_{\bar\sigma}^{\dagger}\;,
  \nonumber\\
  \I \frac{\D}{\D t} X^{\bar\sigma 2}(t) &=&
  [X^{\bar\sigma 2},\tilde{H}_{\mathrm{eff}}]
 = (U-\mu) X^{\bar\sigma 2}
  \label{eq3.8}\\
&+& \sigma V(X^{22}+ X^{\bar\sigma \bar\sigma}) \xi_{\sigma}
{-}\sigma VX^{\bar\sigma \sigma} \xi_{\bar\sigma} {-}
VX^{02}\xi_{\bar\sigma}^{\dagger}\;.
  \nonumber
\end{eqnarray}

In the Green's functions of higher order we shall separate the irreducible
parts using the method developed in \cite{Tserkovnikov,PlakidaPLA}.
Proceeding from the equations of motion (\ref{eq3.8}) we express
derivatives $\I\, \D X^{0\sigma(\bar\sigma 2)}/\D t$ as a sum of regular
(projected on the subspace formed by operators $X^{0\sigma}$,
$X^{\bar\sigma 2}$) and irregular parts. The latter ones describe an
inelastic quasiparticle scattering. We obtain
\begin{eqnarray}
  [X^{0\sigma},\tilde{H}_{\mathrm{eff}}] &=& -\mu X^{0\sigma} +
  \alpha_{1}^{0\sigma} X^{0\sigma} +
  \alpha_{2}^{0\sigma} X^{\bar\sigma 2} + Z^{0\sigma}\;,
  \nonumber\\
  {}[X^{\bar\sigma 2}, \tilde{H}_{\mathrm{eff}}] &=&(U-\mu) X^{\bar\sigma 2} +
  \alpha_{1}^{\bar\sigma 2} X^{0\sigma} \nonumber\\
  &+&
  \alpha_{2}^{\bar\sigma 2}X^{\bar\sigma 2} +
  Z^{\bar\sigma 2}\;.
  \label{eq3.10}
\end{eqnarray}

Operators $Z^{0\sigma}$ and $Z^{\bar\sigma 2}$ are defined as orthogonal
ones to operators from the basic subspace:
\begin{eqnarray}
  \langle\{ Z^{0\sigma(\bar\sigma 2)}, X^{\sigma 0}\}\rangle &=&0\;,
  \nonumber\\
  \langle\{ Z^{0\sigma(\bar\sigma 2)}, X^{2\bar\sigma}\}\rangle &=&0\;.
  \label{eq3.11}
\end{eqnarray}
These equations determine the coefficients
$\alpha_{i}^{0\sigma(\bar\sigma2)}$.

Using the described procedure we come to the expressions
\begin{eqnarray}
  Z^{0\sigma} &=& V\overline{(X^{00}+X^{\sigma\sigma})\xi_{\sigma}} +
  V\overline{X^{\bar\sigma \sigma}\xi_{\bar\sigma}} + \sigma
  V\overline{X^{02}\xi_{\bar\sigma}^{\dagger}}\;, \nonumber\\
  Z^{\bar\sigma 2} &=& \sigma
  V\overline{(X^{22}+X^{\bar\sigma \bar\sigma})\xi_{\sigma}} -
  \sigma V\overline{X^{\bar\sigma
  \sigma}\xi_{\bar\sigma}}\nonumber\\
  &-& V\overline{X^{02}\xi_{\bar\sigma}^{\dagger}}\;,
  \label{eq3.12}
\end{eqnarray}
where
\begin{eqnarray}
  &&\overline{(X^{00}+X^{\sigma\sigma})\xi_{\sigma}} = (X^{00} +
  X^{\sigma\sigma}) \xi_{\sigma}\;, \nonumber\\
  &&\overline{(X^{22}+X^{\bar\sigma \bar\sigma})\xi_{\sigma}} = (X^{22} +
  X^{\bar\sigma \bar\sigma})\xi_{\sigma}\;,  \nonumber\\
  &&\overline{X^{\bar\sigma \sigma} \xi_{\bar\sigma}} = X^{\bar\sigma \sigma}
  \xi_{\bar\sigma} - \frac{1}{A_{0\sigma}}\langle\xi_{\bar\sigma}
  X^{\bar\sigma 0}\rangle X^{0\sigma} \nonumber\\
  && \qquad\quad - \frac{1}{A_{2\bar\sigma}} \langle
  X^{2\sigma} \xi_{\bar\sigma}\rangle X^{\bar\sigma 2}\;, \nonumber\\
  &&\overline{X^{02}\xi_{\bar\sigma}^{\dagger}} = X^{02} \xi_{\bar\sigma}^{\dagger} -
  \frac{1}{A_{0\sigma}} \langle X^{\sigma 2}\xi_{\bar\sigma}^{\dagger}\rangle
  X^{0\sigma} \nonumber\\
  && \qquad\quad - \frac{1}{A_{2\bar\sigma}} \langle \xi_{\bar\sigma}^{\dagger}
  X^{0\bar\sigma}\rangle X^{\bar\sigma 2}
  \label{eq3.13}
\end{eqnarray}
and $A_{pq} = \langle X^{pp} + X^{qq}\rangle$;
$A_{0\sigma}=1-n_{\bar\sigma}$, $A_{\bar\sigma 2}=n_{\bar\sigma}$.

Here
\begin{eqnarray}
  \alpha_{1}^{0\sigma} &=&- \sigma\alpha_{1}^{\bar\sigma 2} =
  \frac{V}{A_{0\sigma}} \varphi_{\sigma}\;,
   \nonumber\\
  \alpha_{2}^{0\sigma} &=&- \sigma\alpha_{2}^{\bar\sigma 2} =
  -\frac{V}{A_{2\bar\sigma}} \sigma \varphi_{\sigma}\;, \nonumber\\
  \varphi_{\sigma} &=& \langle\xi_{\bar\sigma} X^{\bar\sigma 0}\rangle +
  \sigma\langle X^{\sigma 2} \xi_{\bar\sigma}^{\dagger}\rangle \;.
  \label{eq3.14}
\end{eqnarray}
(we put $\varphi_{\sigma}=\varphi_{\sigma}^{*}$).

The equations for the first two functions in (\ref{eq3.7}) have in this
case the form
\begin{eqnarray}
&&  \left(\begin{array}{cc}
  \omega-a_{\sigma} & \sigma\frac{V}{A_{2\bar\sigma}}\varphi_{\sigma} \\
  \sigma\frac{V}{A_{0\sigma}}\varphi_{\sigma} & \omega - b_{\sigma}
  \end{array} \right)
  \left(\begin{array}{c}
  \langle\langle X^{0\sigma} | X^{\sigma 0}\rangle\rangle \\
  \langle\langle X^{\bar\sigma 2} | X^{\sigma 0} \rangle\rangle
  \end{array}  \right) \nonumber\\
&& \qquad\qquad\qquad  =
  \left(\begin{array}{c}
  \frac{A_{0\sigma}}{2\pi} +\langle\langle Z^{0\sigma}|X^{\sigma
  0}\rangle\rangle \\
  \langle\langle Z^{\bar\sigma 2}|X^{\sigma 0}\rangle\rangle
  \end{array} \right)\;,
  \label{eq3.15}
\end{eqnarray}
where the notations
\begin{equation}
  a_{\sigma} = -\mu + \frac{V}{A_{0\sigma}} \varphi_{\sigma}\;, \qquad
  b_{\sigma} = U-\mu+\frac{V}{A_{2\bar\sigma}} \varphi_{\sigma}
  \label{eq3.16}
\end{equation}
are used (a similar set of equations can be written for functions
$\langle\langle Z^{0\sigma(\bar\sigma  2)}
|X^{2\bar\sigma}\rangle\rangle$).

An equation for the Green's function $\langle\langle Z^{0\sigma(\bar\sigma
2)} |X^{\sigma 0}\rangle\rangle$ (as well as for the function
$\langle\langle Z^{0\sigma(\bar\sigma 2)} |X^{2\bar\sigma}\rangle\rangle$)
can be obtained by means of the differentiation with respect to the second
time argument. Applying the similar procedure of separation of the
irregular parts we obtain the expression
\begin{equation}
  \hat{G} = \hat{G}_{0}+\hat{G}_{0}\hat{P}_{\sigma}\hat{G}_{0}\;,
  \label{eq3.19}
\end{equation}
where the matrix Green's function
\begin{equation}
  \hat{G} = 2\pi \left(\begin{array}{cc}
  \langle\langle X^{0\sigma}|X^{\sigma 0}\rangle\rangle & \langle\langle
  X^{0\sigma} | X^{2\bar\sigma} \rangle\rangle \\
  \langle\langle X^{\bar\sigma 2}|X^{\sigma 0}\rangle\rangle & \langle\langle
  X^{\bar\sigma 2} | X^{2\bar\sigma} \rangle\rangle \end{array}\right)
  \label{eq3.18}
\end{equation}
is introduced. $\hat{G}_{0}$ is a nonperturbed Green's function
\begin{equation}
  \hat{G}_{0} = \frac{1}{D_{\sigma}} \left(\begin{array}{cc}
  \omega-b_{\sigma} & -\sigma\frac{V}{A_{2\bar\sigma}}\varphi_{\sigma} \\
  -\sigma\frac{V}{A_{0\sigma}} \varphi_{\sigma} & \omega-a_{\sigma}
  \end{array} \right) \left(\begin{array}{cc}
  A_{0\sigma} & 0 \\
  0 & A_{2\bar\sigma} \end{array} \right)\;,
  \label{eq3.20}
\end{equation}
where
\begin{equation}
  D_{\sigma} = (\omega-a_{\sigma})(\omega-b_{\sigma}) -
  \frac{V^{2}}{A_{0\sigma}A_{2\bar\sigma}}\varphi_{\sigma}^{2}
  \label{eq3.21}
\end{equation}
and
\begin{eqnarray}
  \hat{P}_{\sigma}&=&2\pi \left( \begin{array}{cc}
  A_{0\sigma}^{-1} & 0 \\
  0 & A_{2\bar\sigma}^{-1} \end{array}\right)
  \left(\begin{array}{cc}
  \langle\langle Z^{0\sigma} | Z^{\sigma 0}\rangle\rangle  &
  \langle\langle Z^{0\sigma} | Z^{2\bar\sigma} \rangle\rangle \\
  \langle\langle Z^{\bar\sigma  2} | Z^{\sigma 0}\rangle\rangle &
  \langle\langle Z^{\bar\sigma  2} | Z^{2\bar\sigma} \rangle\rangle
  \end{array} \right) \nonumber\\
&\times&  \left(\begin{array}{cc}
  A_{0\sigma}^{-1} & 0 \\
  0 & A_{2\bar\sigma}^{-1} \end{array} \right)
  \label{eq3.22}
\end{eqnarray}
has the meaning of a scattering matrix. Being expressed in terms of
irreducible Green's functions it contains the scattering corrections of
the second and higher order in powers of $V$. The separation in $\hat{P}$
of the irreducible, with respect to $V$, parts enables us to obtain a mass
operator $\hat{M}$
\begin{eqnarray}
  \hat{P}_{\sigma} &=& \hat{M}_{\sigma} +
  \hat{M}_{\sigma}\hat{G}_{0}\hat{M}_{\sigma} +
  \hat{M}_{\sigma}\hat{G}_{0}\hat{M}_{\sigma}\hat{G}_{0}\hat{M}_{\sigma} +
  \ldots\,, \nonumber\\
  \hat{M}_{\sigma} &=& \hat{P}_{\sigma}|_{\mathrm{ir}}\;.
  \label{eq3.23}
\end{eqnarray}

In this case the relation (\ref{eq3.19}) can be transformed into the Dyson
equation
\begin{equation}
  \hat{G} = \hat{G}_{0} + \hat{G}_{0} \hat{M}_{\sigma} \hat{G}
  \label{eq3.24}
\end{equation}
with the solution
\begin{equation}
  \hat{G}=(1-\hat{G}_{0}\hat{M}_{\sigma})^{-1} \hat{G}_{0}\;,
  \label{eq3.25}
\end{equation}
which provides a final expression for the Green's function (\ref{eq3.18}).

\section{Different-time decoupling of irre\-du\-c\-ible Green's functions}

We will restrict ourselves hereafter to the simple approximation in
calculating the mass operator $\hat{P}$, taking into account the
scattering processes of the second order in $V$. In this case
\begin{equation}
  \hat{M}_{\sigma} = \hat{P}_{\sigma}^{(0)}\;,
  \label{eq4.1}
\end{equation}
where the irreducible Green's functions are calculated without allowance
for correlation between electron transitions on the given site and
environment. It corresponds to the procedure of the different-time
decoupling \cite{PlakidaBog}, which means in our case an independent
averaging of the products of $X$ and $\xi$ operators.

Let us illustrate this approximation with some examples.

\bigskip

1. The Green's function \newline
$\langle\langle
\overline{\vphantom{\xi_{\sigma}^{\dagger}}(X^{00}+X^{\sigma\sigma})\xi_{\sigma}}
| \overline{\xi_{\sigma}^{\dagger}
(X^{00}+X^{\sigma\sigma})}\rangle\rangle_{\omega} \equiv
I_{1}(\omega)$.

According to the spectral theorem we have
\begin{eqnarray}
  I_{1}(\omega) &=& \frac{1}{2\pi}\! \int\limits_{-\infty}^{+\infty}\! \frac{\D\omega'}
  {\omega-\omega'} (\E^{\beta\omega'}+1) \!\int\limits_{-\infty}^{+\infty}\!
  \frac{\D t}{2\pi} \E^{-\I\omega't} \langle \xi_{\sigma}^{\dagger}(t) \nonumber\\
&\times& (X^{00}+
  X^{\sigma\sigma})_{t} (X^{00}+X^{\sigma\sigma})\xi_{\sigma}\rangle^{\mathrm{ir}}\;.
  \label{eq4.2}
\end{eqnarray}
Due to the different-time decoupling
\begin{eqnarray}
&&  \langle \xi_{\sigma}^{\dagger}(t)
(X^{00}+X^{\sigma\sigma})_{t}
  (X^{00}+X^{\sigma\sigma}) \xi_{\sigma}\rangle^{\mathrm{ir}}
  \nonumber\\
&& \qquad  \approx
  \langle (X^{00} + X^{\sigma\sigma})_{t} (X^{00}+X^{\sigma\sigma})\rangle
  \langle\xi_{\sigma}^{\dagger}(t)\xi_{\sigma}\rangle \;.
  \label{eq4.3}
\end{eqnarray}
We will take the first of these correlators in a zero approximation
\begin{eqnarray}
&&  \langle
(X^{00}+X^{\sigma\sigma})_{t}(X^{00}+X^{\sigma\sigma})\rangle
\nonumber\\
&& \qquad  \approx \langle (X^{00}+X^{\sigma\sigma})^{2}\rangle
=A_{0\sigma}\;,
  \label{eq4.4}
\end{eqnarray}
and substitution of (\ref{eq4.3}) into (\ref{eq4.2}) leads in this case to
the result
\begin{equation}
  I_{1}(\omega) = A_{0\sigma} \langle\langle \xi_{\sigma}
  |\xi_{\sigma}^{\dagger}\rangle\rangle_{\omega}=
  \frac{A_{0\sigma}}{2\pi V^2} J_{\sigma}(\omega)\;.
  \label{eq4.5}
\end{equation}

\bigskip

2. The Green's function $\langle\langle
\overline{\vphantom{\xi_{\bar\sigma}^{\dagger}}X^{\bar\sigma
\sigma} \xi_{\bar\sigma}} | \overline{\xi_{\bar\sigma}^{\dagger}
X^{\sigma\bar\sigma}}\rangle\rangle_{\omega}\equiv I_{2}(\omega)$.

The representation of the $I_{2}(\omega)$ function in the form analogous
to (\ref{eq4.2}) leads to the time correlation function $\langle
\xi_{\bar\sigma}^{\dagger}(t) X^{\sigma\bar\sigma}(t)
X^{\bar\sigma\sigma}\xi_{\bar\sigma}\rangle^{\mathrm{ir}}$ that can be
approximated as
\begin{eqnarray}
  \langle \xi_{\bar\sigma}^{\dagger}(t) X^{\sigma\bar\sigma} (t)
X^{\bar\sigma \sigma}
  \xi_{\bar\sigma}\rangle^{\mathrm{ir}} &\approx & \langle X^{\sigma\bar\sigma}(t)
  X^{\bar\sigma \sigma}\rangle \langle
  \xi_{\bar\sigma}^{\dagger}(t)\xi_{\bar\sigma}\rangle \nonumber\\
  &\approx & \langle X^{\sigma\sigma}\rangle \langle
  \xi_{\bar\sigma}^{\dagger}(t)\xi_{\bar\sigma}\rangle\;.
  \label{eq4.6}
\end{eqnarray}
In this case
\begin{equation}
  I_{2}(\omega) = \langle X^{\sigma\sigma}\rangle \langle\langle \xi_{\bar\sigma}
  | \xi_{\bar\sigma}^{\dagger}\rangle\rangle_{\omega}
  =\frac{\langle X^{\sigma\sigma}\rangle}{2\pi V^2} J_{\bar\sigma}(\omega)\;.
  \label{eq4.8}
\end{equation}

\bigskip

3. The Green's function $\langle\langle
\overline{X^{02}\xi_{\bar\sigma}^{\dagger}} |
\overline{\vphantom{\xi_{\bar\sigma}^{\dagger}}\xi_{\bar\sigma}
X^{20}}\rangle\rangle_{\omega}\equiv I_{3}(\omega)$.

The corresponding time correlation function is decoupled as
\begin{eqnarray}
&&  \langle \xi_{\bar\sigma}(t) X^{20}(t) X^{02}
\xi_{\bar\sigma}^{\dagger}\rangle^{\mathrm{ir}}
\nonumber\\
&& \qquad  \approx \exp [\I (U-2\mu)t]\langle X^{22}\rangle
  \langle \xi_{\bar\sigma}(t)\xi_{\bar\sigma}^{\dagger} \rangle \;.
  \label{eq4.9}
\end{eqnarray}
Using this expression we obtain
\begin{eqnarray}
  I_{3}(\omega) &=& \frac{1}{2} \langle X^{00} {+} X^{22}\rangle
  \langle\langle \xi_{\bar\sigma}^{\dagger} | \xi_{\bar\sigma}
  \rangle\rangle_{\omega{+}2\mu{-}U}
  \nonumber \\
&+& \langle X^{00} {-}
  X^{22}\rangle \frac{1}{2\pi} \int\limits_{{-}\infty}^{+\infty}
  \frac{\D\omega'}{\omega {+}2\mu{-}U{-}\omega'} \nonumber\\
&\times& {\tanh}\frac{\beta\omega'}{2}
  [{-}2 \Im \langle\langle \xi_{\bar\sigma}^{\dagger} |
  \xi_{\bar\sigma}\rangle\rangle_{\omega'{+}\I \varepsilon}]\;.
  \label{eq4.11}
\end{eqnarray}
Let us mention that at the half-filling of electron states (when $n=1$,
$\langle X^{00}\rangle = \langle X^{22}\rangle$)
\begin{eqnarray}
  I_{3}(\omega) &=&- \langle X^{22}\rangle \langle\langle \xi_{\bar\sigma}
  |\xi_{\bar\sigma}^{\dagger}\rangle\rangle_{U-2\mu-\omega}
  \nonumber\\
&=& -\frac{\langle X^{22}\rangle}{2\pi V^2}
J_{\bar\sigma}(U-2\mu-\omega)\;.
  \label{eq4.12}
\end{eqnarray}

\bigskip

Following  the described procedure and taking into account the relation
(\ref{eq3.4}) we will come to the following expressions for irreducible
Green's functions:
\begin{eqnarray}
  \langle\langle Z^{0\sigma} | Z^{\sigma 0}\rangle\rangle_{\omega} &=&
  A_{0\sigma} J_{\sigma} (\omega) - R_{\sigma}(\omega)\;, \nonumber\\
  \langle\langle Z^{\bar\sigma 2} |Z^{2\bar\sigma}\rangle\rangle_{\omega} &=&
  A_{2\bar\sigma} J_{\sigma}(\omega) - R_{\sigma}(\omega)\;, \nonumber\\
  \langle\langle Z^{0\sigma} | Z^{2\bar\sigma} \rangle\rangle_{\omega} &=&
  \langle\langle Z^{\bar\sigma 2} | Z^{\sigma 0}\rangle\rangle_{\omega} =
  R_{\sigma}(\omega)\;,
  \label{eq4.13}
\end{eqnarray}
where
\begin{eqnarray}
  \lefteqn{R_{\sigma}(\omega) =
  \frac{1}{2} \langle X^{00} + X^{22}\rangle J_{\bar\sigma} (U-2\mu-\omega)
  -\langle X^{\sigma\sigma}\rangle J_{\bar\sigma}(\omega)
  }
  &&
  \nonumber\\
  &&{}- \langle X^{00} - X^{22}\rangle \!\!\!\int\limits_{-\infty}^{+\infty}\!\!
  \frac{\D\omega'}{\pi}
  \frac{\Im J_{\bar\sigma}(-\omega'-\I 0^+)}{\omega+2\mu-U-\omega'}
  \tanh \frac{\beta\omega'}2\;.
  \label{eq4.14}
\end{eqnarray}

\section{Basic set of equations}

Using the results obtained in the previous section we can write the
expressions for mass operator components $M_{\sigma,\alpha\beta}$. On the
basis of relation
\begin{equation}
  \Xi_{\sigma}^{-1}(\omega) = \Bigl(\sum_{\alpha\beta}
  G_{\alpha\beta}^{\sigma}(\omega)\Bigr)^{-1} +J_{\sigma}(\omega)\;,
  \label{eq5.4}
\end{equation}
[which follows from (\ref{eq2.11})] and formula (\ref{eq3.25}) it is
possible to determine the single-site self-energy part. We obtain
\begin{eqnarray}
  \Xi_{\sigma}(\omega) &=& \left[\omega-\varepsilon_{+}A_{0\sigma}
  - \varepsilon_{-}A_{2\bar\sigma} -
  \tilde{\Omega}_{\sigma}(\omega)\right] \nonumber\\
&\times&
  \left[(\omega-\varepsilon_{+})(\omega-\varepsilon_{-}) -
  \omega\tilde{\Omega}_{\sigma}(\omega) \right. \nonumber\\
&+& \left.
(\varepsilon_{+}A_{2\bar\sigma}+\varepsilon_{-}A_{0\sigma})
  \tilde{\Omega}_{\sigma}(\omega)\right]^{-1}\;,
\label{eq5.5}
\end{eqnarray}
where $\varepsilon_{+}=U-\mu$, $\varepsilon_{-}=-\mu$ and
\begin{eqnarray}
  \tilde{\Omega}_{\sigma}(\omega) &=& \Omega_{\sigma}(\omega) +
  \frac{V\varphi_{\sigma}}{A_{0\sigma}A_{2\bar\sigma}}\;, \nonumber\\
  \Omega_{\sigma}(\omega) &=& J_{\sigma}(\omega) - \frac{R_{\sigma}(\omega)}
  {A_{0\sigma}A_{2\bar\sigma}}\;.
  \label{eq5.6}
\end{eqnarray}
It should be mentioned that formula (\ref{eq5.5}) can be also represented
in the form
\begin{eqnarray}
  \Xi_{\sigma}^{-1}(\omega) &=&
  \left[\frac{A_{0\sigma}}{\omega-\varepsilon_{-} -\tilde{\Omega}_{\sigma}
  (\omega)} + \frac{A_{2\bar\sigma}} {\omega-\varepsilon_{+} -
  \tilde{\Omega}_{\sigma}(\omega)}\right]^{-1} \nonumber\\
  &+& \tilde{\Omega}_{\sigma}
  (\omega)\;.
  \label{eq5.7}
\end{eqnarray}

The relation (\ref{eq5.5}) together with (\ref{eq2.8}), (\ref{eq2.11}) and
(\ref{eq2.12}) creates a set of equations for the coherent potential
$J_{\sigma}(\omega)$, self-energy part $\Xi_{\sigma}(\omega)$ and Green's
functions $G_{ii,\sigma}(\omega)$ and $G_{\vec{k}}^{\sigma}(\omega)$.

It should be noted that the parameter $\varphi_{\sigma}$, which is
expressed in terms of average values of the products of $X$ and $\xi$
operators (formula (\ref{eq3.14})), is a functional of the potential
$J_{\sigma}(\omega)$. According to the spectral theorem
\begin{eqnarray}
  V\langle X^{\sigma 0 (2\bar\sigma)}\xi_{\sigma}\rangle &=&
  \I \int\limits_{-\infty}^{+\infty} \frac{\D\omega}{\E^{\beta\omega}+1}
  \left[ V
  \langle\langle \xi_{\sigma}|X^{\sigma
  0(2\bar\sigma)}\rangle\rangle_{\omega+\I 0^+}  \right. \nonumber\\
&-& \left. V\langle\langle
  \xi_{\sigma}|X^{\sigma 0(2\bar\sigma)}\rangle\rangle_{\omega-\I 0^+}
  \right]\;.
  \label{eq5.8}
\end{eqnarray}

On the other hand, using the linearized equation of motion (\ref{eq3.10})
and neglecting the irreducible parts, we can obtain the following set of
equations

\begin{eqnarray}
&&  V\langle\langle \xi_{\sigma} | X^{\sigma 0}\rangle\rangle
  \left(\omega-\varepsilon_{-} - \frac{V}{A_{0\sigma}}\varphi_{\sigma}\right)
\nonumber\\
&& \quad  + V^{2}\sigma\frac{\varphi_{\sigma}}{A_{2\bar\sigma}}
\langle\langle
  \xi_{\sigma} |X^{2\bar\sigma}\rangle\rangle = \frac{A_{0\sigma}}{2\pi}
  J_{\sigma}\;,   \nonumber\\
&&  V^{2}\sigma\frac{\varphi_{\sigma}}{A_{0\sigma}} \langle\langle
\xi_{\sigma}
  | X^{\sigma 0}\rangle\rangle
 \label{eq5.9} \\
&& \quad + V\langle\langle \xi_{\sigma} |
  X^{2\bar\sigma}\rangle\rangle \left(\omega-\varepsilon_{+} -
  \frac{V}{A_{2\bar\sigma}}\varphi_{\sigma}\right) =
  \sigma\frac{A_{2\bar\sigma}}{2\pi} J_{\sigma}\;. \nonumber
\end{eqnarray}
It follows herefrom in the $U\to\infty$ limit
\begin{eqnarray}
  V\varphi_{\sigma} &=& - V\langle X^{\bar\sigma 0}\xi_{\bar\sigma}\rangle
  =-\frac{1}{2\pi} \int\limits_{-\infty}^{+\infty}
  \frac{\D\omega}{\E^{\beta\omega}+1} \nonumber\\
&\times&   \left[ -2 \Im
\frac{A_{0\bar\sigma}J_{\bar\sigma}(\omega)}
  {\omega-\varepsilon_{-}-V\frac{\varphi_{\bar\sigma}}{A_{0\bar\sigma}}}
  \right]_{\omega+\I 0^+}\;,
  \label{eq5.12}
\end{eqnarray}
or, in the Matsubara's representation
\begin{equation}
  V\varphi_{\sigma}= -\frac{1}{\beta} A_{0\bar\sigma} \sum_{\nu}
  \frac{J_{\bar\sigma}(\omega_{\nu})}
  {\I \omega_{\nu}-\varepsilon_{-}-V\frac{\varphi_{\bar\sigma}} {A_{0\bar\sigma}}}\;.
  \label{eq5.13}
\end{equation}
Thus, we obtain a self-consistent equation for the parameter
$\varphi_{\sigma}$.

\section{Some specific cases}

Equations obtained in the previous section form an approximate analytical
scheme of calculating both the single-site and the full electron Green's
function in the framework of DMFT. Let us compare it with the  standard
approximations known from literature which are based on the assumption of
the single-site structure of the electron self-energy. For this purpose we
will consider some specific cases.

\subsection{Hubbard-I approximation ($J_{\sigma}=0$,
$R_{\sigma}=0$, $\varphi_{\sigma}=0$;
$\tilde{\Omega}_{\sigma}=0$)}

It is the simplest approximation; renormalization of energies of atomic
electron transitions is absent and the scattering processes via coherent
potential are not taken into account. The expression for the single-site
self-energy part
\begin{equation}
  \Xi_{\sigma}(\omega) = \frac{A_{0\sigma}}{\omega-\varepsilon_{-}} +
  \frac{A_{2\bar\sigma}} {\omega-\varepsilon_{+}}
  \label{eq6.1}
\end{equation}
corresponds to the Hubbard-I approximation \cite{H-I}. Electron energy
spectrum described by Green's function $G_{k}^{\sigma}(\omega)$ consists
in this case of two Hubbard subbands divided by a gap existing at any
relationship between the values of $U$ and $t$ parameters.

\subsection{Static mean-field approximation ($J_{\sigma}=0$,
$R_{\sigma}=0$;
$\tilde{\Omega}_{\sigma}=B\equiv V\varphi_{\sigma}/A_{0\sigma}A_{2\bar\sigma}$)}

In this case only a self-consistent shift of the energy levels of the
single-site atomic problem is taken  into account. The coherent potential
$J_{\sigma}(\omega)$ is replaced in the expression for $\varphi_{\sigma}$
by the approximate expression
\begin{equation}
  J_{\sigma}(\omega) = \sum_{kj} t_{ik}t_{ij} G_{kj,\sigma}(\omega)
  \label{eq6.2}
\end{equation}
following from (\ref{eq2.13}) when the difference between
$G_{kj,\sigma}^{(i)}$ and $G_{kj,\sigma}$ is neglected.

The expression for the self-energy part
\begin{eqnarray}
  \Xi_{\sigma}^{-1}(\omega) &=& {G_{\sigma}^{a}}^{-1}(\omega) \nonumber\\
&=& \left[
  \frac{A_{0\sigma}} {\omega-\varepsilon_{-}-B} + \frac{A_{2\bar\sigma}}
  {\omega-\varepsilon_{+}-B}\right]^{-1} \! +B\, ,
  \label{eq6.3}
\end{eqnarray}
that can be obtained in this case, corresponds to the summation of the
series
\begin{equation}
  \hat{G} = \hat{g}_{0}\hat{I} - \hat{g}_{0} \hat{W} \hat{g}_{0} \hat{I}+
  \hat{g}_{0} \hat{W} \hat{g}_{0} \hat{W} \hat{g}_{0} \hat{I} - \ldots\;,
  \label{eq6.4}
\end{equation}
where
\begin{eqnarray}
\hat{g}_{0} &=& \left(
  \begin{array}{cc}
  (\omega-\varepsilon_{-})^{-1} & 0 \\
  0 & (\omega-\varepsilon_{+})^{-1} \end{array}
  \right)\;, \nonumber\\
\hat{I} &=& \frac{1}{2\pi} \left(
  \begin{array}{cc}
  A_{0\sigma} & 0 \\
  0 & A_{2\bar\sigma}
  \end{array}
  \right)\;, \nonumber\\
\hat{W} &=& B\left(\begin{array}{cc}
  -A_{2\bar\sigma} & A_{0\sigma} \\
  A_{2\bar\sigma} & -A_{0\sigma}
  \end{array} \right)\;.
  \label{eq6.5}
\end{eqnarray}

A sum of the  diagrams with loop-like inclusions into the line of the
electron single-site Green's function corresponds to this series in a
diagrammatic representation. Such inclusions lead to the renormalization
of energies of the electron levels \cite{StasyukDanyliv}. In particular,
at $U=\infty$
\begin{equation}
  G_{\sigma}^{(a)}(\omega) = \langle\langle X^{0\sigma}|X^{\sigma
  0}\rangle\rangle_{\omega} = \frac{1}{2\pi} \frac{A_{0\sigma}}
  {\omega-\varepsilon_{-}-V \frac{\varphi_{\bar\sigma}} {A_{0\sigma}}}\;.
  \label{eq6.6}
\end{equation}

Energy shift
\begin{equation}
  \Delta\varepsilon_{-} = A_{2\bar\sigma} B = -\frac{1}{A_{0\sigma}}
  \sum_{l} t_{il}\langle X_{l}^{\bar\sigma 0}
  X_{i}^{0\bar\sigma}\rangle
  \label{eq6.7}
\end{equation}
coincides in this case  with the previously obtained one in a number of
papers (see, for example \cite{PotthoffHWN,H_L,PlakidaYuS}) using a more
complicated (in comparison with Hubbard-I approximation) decoupling
procedure in equations for the Green's function
$G_{\vec{k}}^{\sigma}(\omega)$.

\subsection{Hubbard-III approximation ($\varphi_{\sigma}=0$;
$\tilde{\Omega}_{\sigma}=J_{\sigma}- R_{\sigma}/A_{0\sigma}A_{2\bar\sigma}$)}

We can pass on to this approximation neglecting, at first, the
renormalization of the atomic electron levels and, secondly, approximating
\begin{equation}
  \frac{\langle X^{\sigma\sigma}\rangle}{A_{0\sigma}A_{2\bar\sigma}} \to 1\;,
  \qquad \frac{\langle X^{22}\rangle}{A_{0\sigma}A_{2\bar\sigma}}\to 1
  \label{eq6.8}
\end{equation}
in the case of half filling (when $\langle X^{00}\rangle = \langle
X^{22}\rangle$) in the expression (\ref{eq5.6}), that becomes exact only
in the $U\to 0$ limit. Consequently, an effective potential of dynamical
mean field $\tilde{\Omega}_{\sigma}(\omega)$ takes the form
\begin{equation}
  \tilde{\Omega}_{\sigma}(\omega) = J_{\sigma}(\omega) + J_{\bar\sigma}(\omega)
  -J_{\bar\sigma} (U-2\mu-\omega)\;.
  \label{eq6.9}
\end{equation}

It corresponds (together with the expression (\ref{eq5.5}) for the
electron self-energy) to the Hubbard-III approximation \cite{H-III}. A
potential $\tilde{\Omega}_{\sigma}(\omega)$ includes (besides the coherent
potential $J_{\sigma}(\omega)$) the terms which describe a scattering on
the spin and charge fluctuations. Electron energy spectrum consists in
this case of two subbands only at $U>U_{c}$ where the critical value
$U_{c}$ corresponds to the metal--insulator transition.

\subsection{Alloy-analogy (AA) approximation ($R_{\sigma}=0$,
$\varphi_{\sigma}=0$; $\tilde{\Omega}_{\sigma}(\omega) =
J_{\sigma}(\omega)$)}

The scattering processes are taken here into account only via coherent
potential. The single-site Green's function looks like
\begin{equation}
  G_{ii,\sigma}(\omega) = G_{\sigma}^{(a)}(\omega) =\frac{A_{0\sigma}}
  {\omega-\varepsilon_{-}-J_{\sigma}} + \frac{A_{2\bar\sigma}}
  {\omega-\varepsilon_{+}-J_{\sigma}}
  \label{eq6.10}
\end{equation}
in this case. This expression is analogous to the locator function for a
binary alloy \cite{Ehrenreich}. The procedure of the
$G_{\vec{k}}^{\sigma}(\omega)$ function calculation corresponds to the CPA
method. Let us write for this approximation an irreducible, according to
Dyson, self-energy part $\Sigma_{\sigma}= \omega-\varepsilon_{-}
-\Xi_{\sigma}^{-1}$, using the expression (\ref{eq5.5}) at
$\tilde{\Omega}_{\sigma}=J_{\sigma}$:
\begin{equation}
  \Sigma_{\sigma} = A_{2\bar\sigma} U\Biggm/ \left(1-\frac{A_{0\sigma} U}
  {\omega+\mu-J_{\sigma}}\right)\;.
  \label{eq6.11}
\end{equation}
Or, after excluding a coherent potential
\begin{equation}
  \Sigma_{\sigma}(\omega) = \frac{A_{2\bar\sigma} U}{1-G_{ii}^{\sigma}(\omega)
  (U-\Sigma_{\sigma}(\omega))}\;.
  \label{eq6.12}
\end{equation}

This equation corresponds to the alloy-analogy (AA) approximation
\cite{PotthoffHWN}.

\subsection{Modified alloy-analogy (MAA) approximation
($R_{\sigma}=0$; $\tilde{\Omega}_{\sigma}=J_{\sigma}+
V\varphi_{\sigma}/A_{0\sigma}A_{2,\sigma} \equiv J_{\sigma}+B$)}

An AA-approach is supplemented here by the inclusion of renormalization of
single-site electron levels. It can now be obtained  that
\begin{equation}
  \Sigma_{\sigma} = A_{2\bar\sigma}U \Biggm/
  \left(1-\frac{A_{0\sigma}U}{\omega-\varepsilon_{-}-\tilde{\Omega}}\right)\;.
  \label{eq6.13}
\end{equation}
This relation can be transformed into the equation
\begin{equation}
  \Sigma_{\sigma} (\omega) = A_{2\bar\sigma}U \Biggm/
  \left(1-
  \frac{G_{ii,\sigma}(U-\Sigma_{\sigma})} {1-G_{ii,\sigma}B}\right)
  \label{eq6.14}
\end{equation}
known in the so-called Modified AA-approach \cite{PotthoffHWN,Herrmann}.

\bigskip

One can see from the quoted specific cases that the approach developed in
this work includes a number of known approximations giving in addition
their unification and generalization. The proposed scheme is more complete
than Hubbard-III approximation (which in its turn is the most general of
the quoted ones) and differs from it by the allowance for a
self-consistent renormalization (due to the static internal field) of the
local energy spectrum as well as by the modification of the potential
$\tilde{\Omega}_{\sigma}$ constituent parts to a more elaborated inclusion
of the magnon and charge fluctuation scattering processes. Participation
of Bose-particles in such a scattering is taken into account in our scheme
in a more consistent way.

Quantitative changes in the electron spectrum (in particular, in the
electron density of states) and then in the thermodynamics of the model,
that might be the consequence of applying the approach suggested herein,
can be the subject of subsequent calculations with the use of numerical
methods.

\section{Simple applications of the method}

Let us demonstrate here the potentialities of the developed approximative
scheme using the examples of two models (the Falicov--Kimball model and
the simplified pseu\-do\-spin--elec\-tron model) which are analytically
solvable in the DMFT approach.

\subsection{Falicov--Kimball model}

The Falicov--Kimball model in its initial version \cite{Falicov} was
proposed for the description of the metal-insulator transformation in
compounds with the transition and rare-earth elements. The itinerant and
localized electrons are included into consideration in the model. The
simplified but sufficiently complete formulation of the FK model was given
in the set of subsequent publications where the Hamiltonian was considered
as a specific case of the Hamiltonian of the Hubbard model on the
assumption that the electron transfer from one lattice site to another one
takes place only in the case of the selected ($\sigma=\uparrow$)
orientation of spins (see for example
\cite{BrandtMielsch,Letfulov98,Letfulov99,Gruber}). Electrons with the
opposite spin orientation ($\sigma=\downarrow$) remain localized. They can
effect the energy of delocalized electrons being a source of scattering.
In this case the Hamiltonian of the model is the following
\cite{BrandtMielsch,Letfulov98,Letfulov99}
\begin{equation}\label{eq7.01}
  H=\sum_i\left(Un_{i\uparrow}n_{i\downarrow}-\mu\sum_{\sigma}n_{i\sigma}\right)
  +\sum_{<ij>}t_{ij}a^{\dag}_{i\uparrow}a_{j\uparrow}\;.
\end{equation}

It should be mentioned that in (\ref{eq7.01}) the itinerant and localized
particles are of the same nature and possess a common chemical potential.
Extension of the model on the case of the motionless particles of
different nature (ions, impurities, spins, etc.) results in the
Hamiltonian of the system of spinless fermions on a lattice which are
moving in the random field given by the variables ($w_j=0,1$)
\cite{Freericks1993,Freericks93b,Freericks99,Freericks00,Freericks02}
\begin{eqnarray}
  H&=&\sum_{ij}t_{ij}c^{\dag}_ic_j+U\sum_ic^{\dag}_ic_iw_i
\nonumber\\
&-& \mu\sum_ic^{\dag}_ic_i+E\sum_iw_i.
\label{eq7.02}
\end{eqnarray}
Here the chemical potentials of electrons ($\mu$) and immobile particles
($-E$) are different. Among various applications the Hamiltonian
(\ref{eq7.02}) describes the conducting electron subsystem of the binary
alloy.

Thermodynamics of the Falicov--Kimball model described by Hamiltonians
(\ref{eq7.01}) and (\ref{eq7.02}) was considered mainly at the fixed
concentration of localized particles $\rho_{\mathrm{ion}} =
\frac1N\sum_iw_i = \mathrm{const}$ ($\rho_{\mathrm{ion}} = \left\langle
n_{i\downarrow}\right\rangle = \mathrm{const}$) and in the regimes of the
given values of the chemical potential $\mu$ or electron concentration:
$\rho_e = \left\langle n_{\uparrow}\right\rangle$ or $n=\sum_{\sigma}
\left\langle n_{\sigma}\right\rangle$
\cite{BrandtMielsch,Freericks1993,Letfulov98,Letfulov99}.

In the DMFT approach the Hamiltonian of the effective single-site problem
looks in the model (\ref{eq7.01}) like
\begin{eqnarray}
  \tilde{H}_{\mathrm{eff}} &=& (U-2\mu) X^{22} - \mu \sum_{\sigma} X^{\sigma\sigma}
\nonumber\\
&+& V \left[\xi_{\uparrow}^{\dagger} (X^{0\uparrow} +
X^{\downarrow2}) + (X^{\uparrow0}+
  X^{2\downarrow})\xi_{\uparrow}\right] \nonumber\\
&+& H_{\xi}\;.
\label{eq7.1}
\end{eqnarray}
In this case, in the equations of motion for $X$-operators
\begin{eqnarray}
  \nonumber
  [X^{0\uparrow},\tilde{H}_{\mathrm{eff}}] &=& -\mu X^{0\uparrow} +
  V(X^{00}+X^{\uparrow\uparrow})\xi_{\uparrow}\;,
  \\{}
  [X^{\downarrow2},\tilde{H}_{\mathrm{eff}}] &=& (U-\mu) X^{\downarrow2} +
  V(X^{22}+X^{\downarrow\downarrow})\xi_{\uparrow}\;.
  \label{eq7.2}
\end{eqnarray}
those terms which are responsible for the scattering with the
participation of Bose-particles (magnons and charge excitations) are
absent. Only the components
\begin{eqnarray}
Z^{0\uparrow} &=& V\overline{(X^{00} +
  X^{\uparrow\uparrow})\xi_{\uparrow}}\;, \nonumber\\
\quad Z^{\downarrow2} &=&
  V\overline{(X^{22}+X^{\downarrow\downarrow})\xi_{\uparrow}}
  \label{eq7.3}
\end{eqnarray}
of irregular parts of the time derivatives of $X$-operators are present.

The corresponding irreducible Green's functions are equal to
\begin{eqnarray}
  2\pi \langle\langle Z^{0\uparrow} | Z^{\uparrow0}\rangle\rangle_{\omega} &=&
  2\pi V^{2} A_{0\uparrow}
  G_{\uparrow} (\omega) = A_{0\uparrow} J_{\uparrow}(\omega)\;,
  \nonumber\\
  2\pi \langle\langle Z^{\downarrow2} | Z^{2\downarrow}\rangle\rangle_{\omega} &=& 2\pi
  V^{2} A_{\downarrow2} G_{\uparrow} (\omega) = A_{\downarrow2} J_{\uparrow}(\omega)\;,
  \nonumber\\
  \langle\langle Z^{0\uparrow} | Z^{2\downarrow} \rangle\rangle_{\omega} &=&
  \langle\langle Z^{\downarrow2} | X^{\uparrow0}\rangle\rangle_{\omega}=0\;.
  \label{eq7.4}
\end{eqnarray}

Using now the formulae (\ref{eq3.22})--(\ref{eq3.24}) we obtain
\begin{eqnarray}
  2\pi G_{\uparrow}^{(a)}(\omega) &=&
  \frac{A_{0\uparrow}}{\omega+\mu-J_{\uparrow}(\omega)}
  \nonumber\\
&+& \frac{A_{\downarrow2}}{\omega+\mu-U-J_{\uparrow}(\omega)}\;.
  \label{eq7.5}
\end{eqnarray}
This expression has the same structure as the Green's function for the
AA-appro\-xi\-ma\-ti\-on and is exact for the Falicov--Kimball model (see,
for example, \cite{BrandtMielsch}). It can be seen directly when we write
the equations of motion for operators $Z^{0\uparrow}$, $Z^{\downarrow2}$
and take into account two obstacles: (i) sums $(X^{00}+X^{\sigma\sigma})$
and $(X^{22}+X^{\bar\sigma\bar\sigma})$ are in this case an integral of
motion and (ii) for the $\xi_{\sigma}$ operator the relation
\begin{equation}\label{eq7.5a}
  [\xi_{\sigma},\tilde{H}_{\mathrm{eff}}] = K_{\sigma}(\omega)\xi_{\sigma} +
  V(X^{0\sigma}+X^{\bar\sigma2})
\end{equation}
takes place in the frequency representation; here the function
$K_{\sigma}(\omega)$ is connected with the Fourier transform of the
Green's function (\ref{eq3.2}) by
\begin{equation}\label{eq7.5b}
  [\omega-K_{\sigma}(\omega)]^{-1}=\mathcal{G}_{\sigma}(\omega)\;.
\end{equation}

It should be mentioned that in the auxiliary Fermi-field approach one can
obtain for the Falikov--Kimball model in the framework of the equation of
motion method an exact expression for the grand canonical potential
$\Omega_{a}$ of the effective single-site problem.

Let us represent the exponential operator $\exp{(-\beta
\tilde{H}_{\mathrm{eff}})}$ with the Hamiltonian (\ref{eq7.1}) in the form
\begin{equation}\label{eq7.5c}
  \E^{-\beta \tilde{H}_{\mathrm{eff}}}=n_{\downarrow} \E^{-\beta(H_1+H_{\xi})}+
  (1-n_{\downarrow}) \E^{-\beta(H_2+H_{\xi})}\;,
\end{equation}
where
\begin{eqnarray}
  && H_1=H_1^0+H_{\mathrm{int}}, \quad H_2=H_2^0+H_{\mathrm{int}}\;,
  \nonumber\\
  && H_1^0=(U-\mu)n_{\uparrow}-\mu, \quad H_2^0=-\mu n_{\uparrow}\;,
  \nonumber\\
  && H_{\mathrm{int}} = V (a_{\sigma}^{\dagger}\xi_{\sigma}
  + \xi_{\sigma}^{\dagger}a_{\sigma})\;.
  \label{eq7.5d}
\end{eqnarray}

By averaging over the $\xi$-field and taking the trace over the variables
$a_{\uparrow}$, $a_{\uparrow}^{\dagger}$ we obtain an operator
\begin{equation}\label{eq7.5e}
  \E^{-\beta \tilde{H}_{\downarrow}}=
  n_{\downarrow}Z_{01}\langle\tilde{\sigma}_{\uparrow}(\beta)\rangle_1^0
  +(1-n_{\downarrow})Z_{02}\langle\tilde{\sigma}_{\uparrow}(\beta)\rangle_2^0\;,
\end{equation}
where $\tilde{\sigma}_{\uparrow}(\beta)$ is a $\sigma$-matrix of the form
presented in (\ref{eq2.9}) with the retarded interaction
$V^2\mathcal{G}_{\uparrow}(\tau-\tau')$; $\langle\ldots\rangle_i^0$ is
statistical average with the Hamiltonian $H_i^0$,
\begin{eqnarray}
  && Z_{01}=\E^{\beta\mu}+\E^{-\beta(U-\mu)}\;,
  \nonumber\\
  && Z_{02}=1+\E^{\beta\mu}\;.
  \label{eq7.5f}
\end{eqnarray}

Operator
\begin{equation}\label{eq7.5g}
  \rho_{\downarrow}=\frac1{\mathcal{Z}_{\mathrm{imp}}}
  \E^{-\beta \tilde{H}_{\downarrow}}
\end{equation}
plays the role of the single site (``impurity") statistical operator for
the electrons with spin $\downarrow$;
\begin{equation}\label{eq7.5h}
  \mathcal{Z}_{\mathrm{imp}} =
  \sum_i Z_{0i}\langle\tilde\sigma_{\uparrow}(\beta)\rangle_i^0\;.
\end{equation}
For the grand canonical potential we obtain an expression
\begin{equation}\label{eq7.5i}
  \Omega_{\mathrm{imp}}=-\Theta\ln\mathcal{Z}_{\mathrm{imp}}=
  -\Theta\ln\left(\sum_i Z_{0i}\,\E^{-\beta Q_i}\right)\;,
\end{equation}
where the notation $\beta
Q_i=-\ln\langle\tilde\sigma_{\uparrow}(\beta)\rangle_i^0$ is used.

It is easy to see that the following relations take place
\begin{eqnarray}
  A_{0\uparrow} &=& \frac{1+\E^{\beta\mu}}{\mathcal{Z}_{\mathrm{imp}}}\E^{-\beta
  Q_2}\; ;
\nonumber\\
  A_{\downarrow2} &=& \frac{\E^{\beta\mu}+\E^{-\beta(U-\mu)}}{\mathcal{Z}_{\mathrm{imp}}}
  \E^{-\beta Q_1}\;.
\label{eq7.5j}
\end{eqnarray}

The quantities $Q_1$ and $Q_2$ can be found by means of differentiation
procedure with respect to the interaction constant $V$. From the one side,
using the relations (\ref{eq7.5i}) and (\ref{eq7.5j}) we get
\begin{equation}\label{eq7.5k}
  \frac{\partial\Omega_{\mathrm{imp}}}{\partial V}=
  A_{\downarrow2}\frac{\partial Q_1}{\partial V}+
  A_{0\uparrow}\frac{\partial Q_2}{\partial V}\;.
\end{equation}
From the other side,
\begin{equation}\label{eq7.5l}
  \frac{\partial\Omega_{\mathrm{imp}}}{\partial V}=
  \left\langle\frac{\partial}{\partial V}\tilde{H}_{\mathrm{eff}}\right\rangle
  =2\left\langle\left(X^{\uparrow0}+X^{2\downarrow}\right)\xi_{\uparrow}\right\rangle
\end{equation}
and according to the spectral theorem for Green's functions
\begin{equation}\label{eq7.5m}
  \frac{\partial\Omega_{\mathrm{imp}}}{\partial V}=
  2\int\limits_{-\infty}^{+\infty}\frac{\D\omega}{\E^{\beta\omega}+1}
  2\Im\langle\langle\xi_{\uparrow}|a_{\uparrow}^{\dagger}\rangle\rangle_{\omega-\I 0^+}\;.
\end{equation}

The function
$\langle\langle\xi_{\uparrow}|a_{\uparrow}^{\dagger}\rangle\rangle_{\omega}$
is calculated with the help of the equation of motion method. Using
(\ref{eq7.5a}) we have
\begin{equation}\label{eq7.5n}
  \langle\langle\xi_{\uparrow}|a_{\uparrow}^{\dagger}\rangle\rangle_{\omega}=
  V\mathcal{G}_{\uparrow}(\omega)
  \langle\langle a_{\uparrow}|a_{\uparrow}^{\dagger}\rangle\rangle_{\omega}
\end{equation}
and finally, after substitution expression (\ref{eq7.5})
\begin{eqnarray}
  \langle\langle\xi_{\uparrow}|a_{\uparrow}^{\dagger}\rangle\rangle_{\omega}&=&
  \frac1{2\pi}\left[A_{0\uparrow}
  \frac{V\mathcal{G}_{\uparrow}(\omega)}{\omega+\mu-V^2\mathcal{G}_{\uparrow}(\omega)}
\right. \nonumber\\
&+& \left. A_{\downarrow2}
  \frac{V\mathcal{G}_{\uparrow}(\omega)}{\omega+\mu-U-V^2\mathcal{G}_{\uparrow}(\omega)}
  \right]\;.
\label{eq7.5o}
\end{eqnarray}

Using this expression in (\ref{eq7.5m}) and comparing the obtained result
with (\ref{eq7.5k}) we get a result
\begin{equation}\label{eq7.5p}
  Q_{1,2}=\frac1{\pi}\int\limits_{-\infty}^{+\infty}\frac{\D \omega}{\E^{\beta\omega}+1}
  \Im\ln\left(1-\frac{J_{\uparrow}(\omega+\I 0^+)}{\omega-\varepsilon_{+,-}+\I 0^+}\right)\;.
\end{equation}
It corresponds to the expression which has a form of a sum over the
Matsubara frequencies
\begin{equation}\label{eq7.5q}
  Q_{1,2}=\frac1{\beta}\sum_n \ln
  \left(1-\frac{J_{\uparrow}(\omega_n)}{\I\omega_n-\varepsilon_{+,-}}\right)
\end{equation}
and was obtained for the first time in \cite{BrandtMielsch}. Formulae
(\ref{eq7.5}) for the Green's function and (\ref{eq7.5i},\ref{eq7.5q}) for
the grand canonical potential of the single site problem were used as a
basic expressions for the consideration of the energy spectrum and
thermodynamics of the FK model in
\cite{BrandtMielsch,Letfulov98,Letfulov99,Freericks99,Freericks00} in the
framework of the DMFT.

\subsection{Simplified pseudospin--electron model}

In recent years the pseudospin--electron model (PEM) has been among the
actively investigated models in the theory of strongly correlated electron
systems. The model appeared in connection with the description of the
anharmonic phenomena in the high-$T_c$ superconductors and in search of
the mechanisms that favor the high values of the transition temperatures
into superconducting state. In addition to the Hubbard type correlation an
interaction with the locally anharmonic lattice vibrations (such as
vibrations connected with the oxygen sublattice ions in the
YBa$_2$Cu$_3$O$_{7-\delta}$ crystals \cite{Mustre,Saiko,Gutmann}) is
included into the model. The corresponding degrees of freedom are
described by the pseudospin variables with $S=1/2$. The Hamiltonian of the
PEM has the form analogous to (\ref{eq1.1}) with
\begin{equation}
  H_{i}=Un_{i\uparrow} n_{i\downarrow} - \mu\sum_{\sigma} n_{i\sigma} +
  g\sum_{\sigma} n_{i\sigma}S_{i}^{z} - hS_{i}^{z} + \Omega S_{i}^{x}
  \label{eq7.6}
\end{equation}
(see \cite{Mueller} as well as \cite{Hirsch,Frick,StasyukPC}). Here $h$ is
internal asymmetry field; $\Omega$ is a parameter of the tunnelling type
splitting.

The model described by the Hamiltonian (\ref{eq7.6}) is more complicated
for consideration than the Hubbard one. An analysis of the energy
spectrum, thermodynamics and charge susceptibility of the model was
performed in \cite{StasyukPC,StasyukFerr} in the case $U=\infty$ using the
generalized random phase approximation (GRPA)
\cite{IzyumovLetfulov,IzyumovPRB}. The single-electron spectrum in this
approach is described in the spirit of the Hubbard-I approximation and is
splitted due to the $gnS^z$ interaction into subbands at the any value of
the coupling constant.

The DMFT method in its standard formulation based on the diagrammatic
expansions for the Matsubara Green's functions and on the written in the
form (\ref{eq2.9}) expression for the interaction with the effective field
(a coherent potential) $J_{\sigma}(\tau-\tau')$ was applied by now only in
the case $U=0$ and $\Omega=0$ \cite{ShvaikaJPS}. In the GRPA such a
simplified pseudospin-electron model was considered in
\cite{StasyukShT,StasyukUJP}. It is possible to solve analytically an
effective single site problem for this case too. Let us illustrate this
with the help of the described above approach.

The effective Hamitlonian for the simplified model is as follows
\begin{eqnarray}
  \tilde{H}_{\mathrm{eff}} &=& -\mu\sum_{\sigma} n_{\sigma} + g\sum_{\sigma} n_{\sigma}
  S^{z} - hS^{z}
\nonumber\\
&+& V\sum_{\sigma} (\xi_{\sigma}^{\dagger} a_{\sigma} +
a_{\sigma}^{\dagger}
  \xi_{\sigma}) + H_{\xi}\;.
  \label{eq7.7}
\end{eqnarray}

Let us write the required single-site Green's function in the form
\begin{equation}
  G_{\sigma}^{(a)} (\omega) =\langle \langle P^{+}a_{\sigma} | P^{+}
  a_{\sigma}^{\dagger}\rangle\rangle_{\omega} + \langle\langle P^{-}a_{\sigma} |
  P^{-} a_{\sigma}^{\dagger}\rangle\rangle_{\omega}\;,
  \label{eq7.8}
\end{equation}
where $P^{\pm}=1/2\pm S^{z}$ are the operators projecting into states with
a given pseudospin orientation.

Following the procedure described in section 3 we consider the equation of
motion.
\begin{equation}
  [P^{\pm} a_{\sigma}, \tilde{H}_{\mathrm{eff}}] = E^{\pm} P^{\pm} a_{\sigma} +
  VP^{\pm} \xi_{\sigma}\;,
  \label{eq7.9}
\end{equation}
where $E^{\pm} =-\mu \pm  g/2$. The irregular part in this case is
\begin{equation}
  Z^{\pm}=V\overline{P^{\pm}\xi_{\sigma}} \equiv VP^{\pm}\xi_{\sigma}\;.
  \label{eq7.10}
\end{equation}

The different-time decoupling gives
\begin{equation}
  \langle\langle \overline{P^{\pm} \xi_{\sigma}} | \overline{P^{\pm}
  \xi_{\sigma}^{\dagger}}\rangle\rangle_{\omega} = \langle P^{\pm} \rangle
  \langle\langle \xi_{\sigma} |\xi_{\sigma}^{\dagger}\rangle \rangle_{\omega}\;.
  \label{eq7.11}
\end{equation}
As a result, we obtain Dyson equation
\begin{equation}
  G_{\sigma}^{\pm}(\omega) = G_{0}^{\pm} + G_{0}^{\pm} M_{\sigma}^{\pm}
  G_{\sigma}^{\pm}
  \label{eq7.12}
\end{equation}
with
\begin{equation}
  G_{0}^{\pm} = \frac{\langle P^{\pm}\rangle} {\omega-E^{\pm}}, \quad
  M^{\pm}_{\sigma(\omega)} = \frac{1} {\langle P^{\pm}\rangle}
  J_{\sigma}(\omega)\;.
  \label{eq7.13}
\end{equation}
It follows herefrom that
\begin{equation}
  G_{\sigma}^{(a)}(\omega) =\frac{\langle P^{+}\rangle}
  {\omega-E^{+}-J_{\sigma}(\omega)} + \frac{\langle P^{-}\rangle} {\omega -
  E^{-}-J_{\sigma} (\omega)}\;.
  \label{eq7.14}
\end{equation}
This expression coincides with the one obtained in the $d=\infty$ limit in
the framework of DMFT \cite{ShvaikaJPS}. The different-time decoupling
(\ref{eq7.11}) is also an exact procedure in this case.

Let us mention that the two-pole structure of the single-site Green's
function leads to the effect of the metal--insulator transition type at
$t\sim g$. In the case $t>g$ the electron spectrum consists of one broad
band while at $t<g$ there appears a gap and the splitting into two
Hubbard-type bands takes place \cite{ShvaikaJPS}. It should be stressed on
this occasion that the Hamiltonian of the simplified PEM corresponds to
the Hamiltonian (\ref{eq7.02}) of the FK model at the replacement
$S^z_i=w_i-\frac12$. There exists however an essential difference in the
regimes of thermodynamical averaging: in the PEM the value of the field
$h$ is fixed (which is an analogue of the chemical potential $-E$ in
(\ref{eq7.02})), but not the mean value of pseudospin. The special
features of the electron and pseudospin subsystems behavior and phase
transitions in PEM in this case are investigated and described in
\cite{ShvaikaJPS,StasyukShT,StasyukUJP}.

\section{Perturbation theory in terms of electron hopping}

Presented in the previous sections an irreducible Green's function
approach allows to construct various approximations for the
single-electron properties, but does not allow in general case to describe
the thermodynamics in a self-consistent way. In this section we present a
different approach based on the Wick's theorem and diagrammatic technique
for the Hubbard operators \cite{ShvaikaPRB}.

We consider the lattice electronic system that can be described by the
following generalized statistical operator:
\begin{eqnarray}
  \hat{\rho} &=& \E^{-\beta \hat{H}_{0}}\hat{\sigma}(\beta)\;,
\nonumber\\
  \hat{\sigma}(\beta) &=& T \!\exp \left\{\!-\!\int\limits_{0}^{\beta} \!\!\D\tau
  \!\int\limits_{0}^{\beta}\!\!\D\tau' \right. \nonumber\\
&\times& \left. \vphantom{\int\limits_{0}^{\beta}}
\sum\limits_{ij\sigma} t_{ij}^{\sigma}
  (\tau-\tau') a_{i\sigma}^{\dag}(\tau) a_{j\sigma}(\tau') \right\}\;,
  \label{eq1}
\end{eqnarray}
where
\begin{equation}
  \hat{H}_{0} = \sum\limits_{i} \hat{H}_{i}
  \label{eq2}
\end{equation}
is a sum of the single-site contributions and for the Hubbard model we
have
\begin{eqnarray}
  & H_{i}=Un_{i\uparrow} n_{i\downarrow} - \mu(n_{i\uparrow} +
  n_{i\downarrow}) - h(n_{i\uparrow} - n_{i\downarrow})\;,
  \nonumber\\
  & t_{ij}^{\sigma} (\tau-\tau') = t_{ij} \delta (\tau-\tau')\;.
  \label{eq3}
\end{eqnarray}
In addition, for the Falicov--Kimball model
\begin{equation}
  t_{ij}^{\sigma}(\tau-\tau') = \left\{\begin{array}{cl}
  t_{ij} \delta(\tau-\tau') & \mbox{ for $\sigma=\uparrow$}\\
  0 & \mbox{ for $\sigma=\downarrow$}
  \end{array}
  \right.\;.
  \label{eq4}
\end{equation}
On the other hand, for the auxiliary single-site problem (\ref{eq2.9}) of
the DMFT we must put
\begin{equation}
  t_{ij}^{\sigma}(\tau-\tau')=\delta_{ij}J_{\sigma}(\tau-\tau')\;.
\end{equation}

It is supposed that we know eigenvalues and eigenstates of the zero-order
Hamiltonian (\ref{eq2}),
\begin{equation}
  H_{i}|i,p\rangle = \lambda_{p} |i,p\rangle
\end{equation}
and one can introduce Hubbard operators in terms of which zero-order
Hamiltonian is diagonal
\begin{equation}\label{Hdiag}
  H_{0}=\sum_{i}\sum_{p} \lambda_{p}\hat{X}_{i}^{pp}\;.
\end{equation}

For the Hubbard model we have four states
$|i,p\rangle=|i,n_{i\uparrow},n_{i\downarrow}\rangle$ (\ref{eq3.6}):
$|i,0\rangle=|i,0,0\rangle$ (empty site), $|i,2\rangle = |i,1,1\rangle$
(double occupied site), $|i,\uparrow\rangle=|i,1,0\rangle$ and
$|i,\downarrow\rangle=|i,0,1\rangle$ (sites with spin-up and spin-down
electrons) with energies
\begin{eqnarray}
\lambda_{0} &=& 0\;, \quad \lambda_{2}=U-2\mu\;, \nonumber\\
\lambda_{\downarrow} &=& h-\mu\;, \quad \lambda_{\uparrow} =
-h-\mu\;.
\label{eq6}
\end{eqnarray}
The connection between the electron operators and the Hubbard operators is
the following:
\begin{equation}
  n_{i\sigma} = X_{i}^{22} + X_{i}^{\sigma\sigma}\;; \quad a_{i\sigma} =
  X_i^{0\sigma} + \sigma X_i^{\bar\sigma 2}\;.
  \label{eq7}
\end{equation}

Our aim is to calculate the grand canonical potential functional
\begin{eqnarray}
  &&\Omega=-\frac{1}{\beta} \ln \Sp\hat{\rho} = \Omega_{0} - \frac{1}{\beta} \ln
  \langle\hat{\sigma}(\beta)\rangle_{0}\;,
  \nonumber\\
  &&\Omega_{0}=-\frac{1}{\beta} \ln \Sp \E^{-\beta H_{0}}\;,
  \label{eq8}
\end{eqnarray}
single-electron Green's functions
\begin{equation}
  G_{ij\sigma}(\tau-\tau') =\langle T a_{i\sigma}^{\dag}(\tau) a_{j\sigma}(\tau')
  \rangle = \beta\frac{\delta \Omega}{\delta t_{ij}^{\sigma}(\tau-\tau')}
  \label{eq9}
\end{equation}
and mean values
\begin{eqnarray}
  n_{\sigma}=\frac{1}{N} \sum_{i} \langle n_{i\sigma} \rangle  =
  -\frac{1}{N} \frac{d\Omega}{d\mu_{\sigma}}\;,
  \nonumber\\
  n=n_{\uparrow}+n_{\downarrow}\;; \quad
  m=n_{\uparrow}-n_{\downarrow}\;,
  \label{eq10}
\end{eqnarray}
where $\mu_{\sigma}=\mu+\sigma h$ is a chemical potential for the
electrons with spin $\sigma$. Here, $\langle\ldots\rangle=\Sp(\ldots
\hat{\rho})/Z$, $Z=\Sp\hat{\rho}$, or in the interaction representation
\begin{equation}
  \langle\ldots\rangle = \frac{1}{\langle \hat{\sigma}(\beta)\rangle_{0}}
  \langle\ldots\hat{\sigma}(\beta)\rangle_{0} =
  \langle\ldots\hat{\sigma}(\beta) \rangle_{0c}\;,
  \label{eq11}
\end{equation}
where $\langle\ldots\rangle_{0}= \Sp (\ldots \E^{-\beta H_{0}})/Z_0$;
$Z_{0}=\Sp \E^{-\beta H_{0}}$.

We expand the scattering matrix $\hat{\sigma}(\beta)$ in (\ref{eq1}) into
the series in terms of electron hopping and for $\langle
\sigma(\beta)\rangle_{0}$ we obtain a series of terms that are products of
the hopping integrals and averages of the electron creation and
annihilation operators or, using (\ref{eq7}), Hubbard operators that will
be calculated with the use of the corresponding Wick's theorem.

Wick's theorem for Hubbard operators was formulated in \cite{Slobodyan}
(see also Ref.~\cite{IzyumovBook} and references therein). For the Hubbard
model we can define four diagonal Hubbard operators $X^{pp}$
($p=0,2,\downarrow,\uparrow$) which are of bosonic type, four annihilation
$X^{0\downarrow}$, $X^{0\uparrow}$, $X^{\uparrow 2}$, $X^{\downarrow 2}$
and four conjugated creation fermionic operators, and two annihilation
$X^{\downarrow\uparrow}$, $X^{02}$ and two conjugated creation bosonic
operators. The algebra of $\hat{X}$ operators is defined by the
multiplication rule
\begin{equation}
  X_{i}^{rs} X_{i}^{pq} = \delta_{sp} X_{i}^{rq}\;,
  \label{eq12}
\end{equation}
the conserving condition
\begin{equation}
  \sum_p X_i^{pp}=1
  \label{eq12p}
\end{equation}
and the commutation relations
\begin{equation}
  [X_{i}^{rs},X_{j}^{pq}]_\pm = \delta_{ij} (\delta_{sp} X_{i}^{rq} \pm
  \delta_{rq} X_{i}^{ps})\;,
  \label{eq13}
\end{equation}
where one must use anticommutator when both operators are of the fermionic
type and commutator in all other cases. So, commutator or anticommutator
of two Hubbard operators is not a $c$ number but a new Hubbard operator.
Then the average of a $T$ products of $X$ operators can be evaluated by
the consecutive  pairing, while taking into account standard permutation
rules for bosonic and fermionic operators, of all off-diagonal Hubbard
operators $X^{pq}$ according to the rule (Wick's theorem)
\begin{eqnarray}
  \stackrel{\unitlength=1em\line(0,-1){.3}\vector(-1,0){1.8}\line(-1,0){1.4}\line(0,-1){.3}\quad}
  {X_{i}^{rs}(\tau_{1}) X_{0}^{pq}}(\tau)
  &=& - \delta_{0i} g_{pq} (\tau-\tau_{1}) \nonumber\\
&\times&  [X_{i}^{rs}(\tau_{1}), X_{i}^{pq} (\tau_{1})]_\pm
  \label{eq14}
\end{eqnarray}
until we get the product of the diagonal Hubbard operators only. Here we
introduce the zero-order Green's function
\begin{eqnarray}
&&  g_{pq}(\tau-\tau_{1}) =
  \frac1{\beta}\sum_{\nu} g_{pq}(\omega_{\nu}) \E^{\I\omega_{\nu}(\tau-\tau_1)}
\nonumber\\
&& \quad = \E^{(\tau-\tau_{1})\lambda_{pq}}
  \left\{\begin{array}{ll}
  \pm n_\pm(\lambda_{pq}) & \quad\tau > \tau_{1} \\
  \pm n_\pm(\lambda_{pq})-1 & \quad\tau < \tau_{1}
  \end{array} \right.\;,
\label{eq15}
\end{eqnarray}
where $\lambda_{pq}=\lambda_{p}-\lambda_{q}$ and $n_\pm(\lambda) =
\left(\E^{\beta\lambda}\pm1\right)^{-1}$, and its Fourier  transform is
equal
\begin{equation}
  g_{pq}(\omega_{\nu}) = \frac{1}{\I\omega_{\nu}-\lambda_{pq}}\;.
  \label{eq16}
\end{equation}

Applying such pairing procedure to the expansion of
$\langle\hat{\sigma}(\beta)\rangle_{0}$ we get the following diagrammatic
representation:
\begin{eqnarray}
  \left\langle\hat\sigma(\beta)\right\rangle_0=\left\langle
  \exp\Biggl\{
  \right.\!\!\!
  \raisebox{-13pt}[16pt][13pt]{\includegraphics[scale=0.7]{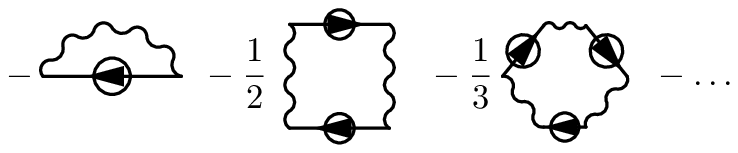}}
  \label{eq17}\\
  \raisebox{-12pt}[15pt][12pt]{\includegraphics[scale=0.7]{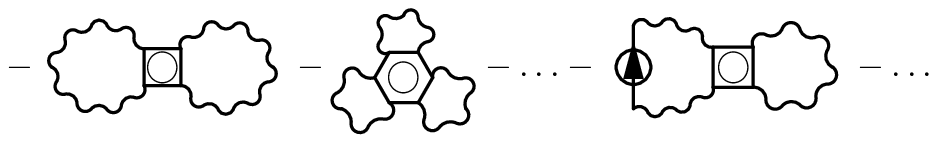}}
  \left.
  \Biggr\}\right\rangle_0\;,
  \nonumber
\end{eqnarray}
where arrows denote the zero-order Green's functions (\ref{eq16}), wavy
lines denote hopping integrals and $\square$, \ldots\ stay for some
complicated ``$n$ vertices'', which for such type perturbation expansion
are an irreducible many-particle single-site Green's functions calculated
with the single-site Hamiltonian (\ref{Hdiag}). Each vertex (Green's
function) is multiplied by a diagonal Hubbard operator denoted by a circle
and one gets an expression with averages of the products of diagonal
Hubbard operators.

For the Falicov--Kimball model expression (\ref{eq17}) reduces and
contains only single loop contributions
\begin{equation}\label{SigmaFK}
  \left\langle\hat\sigma(\beta)\right\rangle_0{=}\left\langle\!
  \exp\Biggl\{
  \raisebox{-13pt}[16pt][13pt]{\includegraphics[scale=0.7]{StasD01.eps}}
  \Biggr\}\right\rangle_0\!,
  \label{eq18}
\end{equation}
where
\[
  \raisebox{-3pt}[6pt][3pt]{\includegraphics[scale=0.7]{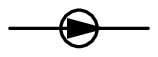}} =\frac{\hat
  P_i^\pm}{\I\omega_{\nu}+\mu^*{\mp}\frac U2}\;;
\]
$\hat{P}_{i}^{+} = \hat{n}_{i\downarrow}$; $\hat{P}^{-} =
1-\hat{n}_{i\downarrow}$, $\mu^*=\mu-U/2$ and by introducing pseudospin
variables $S_{i}^{z}= (\hat{P}_{i}^{+} - \hat{P}_{i}^{-})/2$ one can
transform the Falicov--Kimball model into an Ising-type model with the
effective multisite retarded pseudospin interactions. Expression
(\ref{SigmaFK}) can be obtained from the statistical operator (\ref{eq1})
by performing partial averaging over fermionic variables, which gives an
effective statistical operator for pseudospins (ions).

So, after applying Wick's theorem our problem splits into two problems:
(i) calculation of the irreducible many-particle Green's functions
(vertices) in order to construct expression (\ref{eq17}) and (ii)
calculation of the averages of the products of diagonal Hubbard operators
and summing up the resulting series.

\section{Irreducible many-particle Green's functions}

For the Hubbard model by applying the Wick's theorem for $X$ operators one
gets for two-vertex
\begin{eqnarray}
  \raisebox{-4pt}[9pt][4pt]{\includegraphics{StasD03.eps}}
  &=& g_{\sigma 0} (\omega_{\nu}) (\hat{X}_{i}^{\sigma\sigma} + \hat{X}_{i}^{00})
\nonumber\\
&+& g_{2\bar{\sigma}} (\omega_{\nu}) (\hat{X}_{i}^{22} +
  \hat{X}_{i}^{\bar{\sigma}\bar{\sigma}})\;,
\label{eq19}
\end{eqnarray}
for four-vertex
\begin{eqnarray}
\lefteqn{\unitlength=0.1em
  \strut_{\omega_{\nu+m}\sigma}^{\omega_{\nu}\sigma}
  \raisebox{-7pt}[11pt][7pt]{\includegraphics{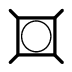}}
  \strut_{\omega_{\nu'+m}\bar\sigma}^{\omega_{\nu'}\bar\sigma}
  =
  \hat\Lambda_{i\sigma\bar\sigma}^{(4)}
  (\omega_{\nu},\omega_{\nu+m},\omega_{\nu'+m},\omega_{\nu'})}
  \label{eq20}\\
  &=& \hat{X}_{i}^{00} g_{\sigma 0}(\omega_{\nu}) g_{\sigma 0} (\omega_{\nu+m})
  \left(U+ U^{2}g_{20} (\omega_{\nu+\nu'+m})\right) \nonumber\\
  && \times g_{\bar{\sigma}0} (\omega_{\nu'})
  g_{\bar{\sigma}0} (\omega_{\nu'+m})
  \nonumber\\
  && +\hat{X}_{i}^{22} g_{2\bar{\sigma}} (\omega_{\nu}) g_{2\bar{\sigma}}
  (\omega_{\nu+m}) \left(U-U^{2} g_{20}(\omega_{\nu+\nu'+m})\right)
\nonumber\\
&& \times g_{2\sigma} (\omega_{\nu'}) g_{2\sigma}
(\omega_{\nu'+m})
  \nonumber\\
  && +\hat{X}_{i}^{\sigma\sigma} g_{\sigma 0} (\omega_{\nu}) g_{\sigma 0}
  (\omega_{\nu+m}) \left(U+U^{2} g_{\sigma\bar{\sigma}} (\omega_{\nu-\nu'})\right)
\nonumber\\
&& \times g_{2\sigma}(\omega_{\nu'}) g_{2\sigma} (\omega_{\nu'+m})
  \nonumber\\
  && +\hat{X}_{i}^{\bar{\sigma}\bar{\sigma}} g_{2\bar{\sigma}} (\omega_{\nu})
  g_{2\bar{\sigma}} (\omega_{\nu+m}) \left(U-U^{2} g_{\sigma\bar{\sigma}}
  (\omega_{\nu-\nu'})\right)
    \nonumber\\
&& \times g_{\bar{\sigma}0} (\omega_{\nu'})
 g_{\bar{\sigma}0}
  (\omega_{\nu'+m})\;,
  \nonumber
\end{eqnarray}
\[
  \hat\Lambda_{i\sigma\sigma}^{(4)}(\omega_{\nu},\omega_{\nu+m},\omega_{\nu'+m},\omega_{\nu'})
  \equiv 0
\]
and so on. Expressions (\ref{eq19}) and (\ref{eq20}) and for the vertices
of higher order possess one significant feature \cite{ShvaikaPRB}. They
decompose into four terms with different diagonal Hubbard operators
$X^{pp}$, which project our single-site problem on certain ``vacuum''
states (subspaces), and zero-order Green's functions, which describe all
possible excitations and scattering processes around these ``vacuum''
states: i.e., creation and annihilation of single electrons and of the
doublon (pair of electrons with opposite spins) for subspaces $p=0$ and
$p=2$ and creation and annihilation of single electrons with appropriate
spin orientation and of the magnon (spin flip) for subspaces $p=\uparrow$
and $p=\downarrow$.

In compact form expressions (\ref{eq19}) and (\ref{eq20}) can be written
as
\begin{equation}
  \raisebox{-4pt}[9pt][4pt]{\includegraphics{StasD03.eps}}
  =\sum_{p} \hat X_{i}^{pp} g_{\sigma(p)}(\omega_{\nu})
  \label{eq21}
\end{equation}
and
\begin{eqnarray}
  \unitlength=0.1em
  \raisebox{-7pt}[11pt][7pt]{\includegraphics{StasD04.eps}}&=&
  \sum_{p} \hat{X}_{i}^{pp}
  g_{\sigma(p)} (\omega_{\nu}) g_{\sigma(p)} (\omega_{\nu+m})
  \widetilde{U}_{\sigma\bar{\sigma}(p)} (\omega_{\nu},\omega_{\nu'}|\omega_{m})
\nonumber \\
&\times& g_{\bar{\sigma}(p)} (\omega_{\nu'}) g_{\bar{\sigma}(p)}
  (\omega_{\nu'+m})\;,
\label{eq22}
\end{eqnarray}
where
\begin{equation}
  g_{\sigma(p)}(\omega_{\nu}) = \left\{ \begin{array}{ll}
  g_{\sigma 0} (\omega_{\nu}) & \mbox{ for $p=0,\sigma$} \\
  g_{2\bar\sigma}(\omega_{\nu}) & \mbox{ for $p=\bar\sigma,2$}
  \end{array} \right.\;.
  \label{eq23}
\end{equation}
Here
\begin{eqnarray}
&&  \widetilde{U}_{\sigma\bar{\sigma}(p)}
(\omega_{\nu},\omega_{\nu'}|\omega_{m})
\nonumber\\
&& \qquad  =
  \left\{ \begin{array} {ll}
  \!\!U\pm U^2 g_{20}(\omega_{\nu+\nu'+m}) & \mbox{for $p=0,2$} \\
  \!\!U\pm U^2 g_{\sigma\bar{\sigma}} (\omega_{\nu-\nu'}) & \mbox{for
  $p=\sigma,\bar{\sigma}$}
  \end{array}
  \right.\;,
  \nonumber\\
&&  \widetilde{U}_{\sigma\bar{\sigma}(p)}
(\omega_{\nu},\omega_{\nu'}|\omega_{m})
 =
  \widetilde{U}_{\bar{\sigma}\sigma(p)} (\omega_{\nu'},\omega_{\nu}|\omega_{m})
\label{eq24}
\end{eqnarray}
is a renormalized Coulomb interaction in the subspaces. In diagrammatic
notations expressions (\ref{eq20}) or (\ref{eq22}) can be represented as
\begin{equation}
  \raisebox{-32pt}[37pt][32pt]{\includegraphics[scale=0.9]{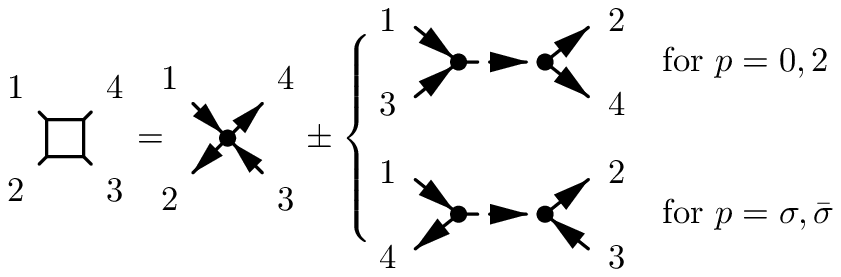}}\;,
  \label{eq25}
\end{equation}
where dots denote Coulomb correlation energy $U=\lambda_{2}+\lambda_{0} -
\lambda_{\uparrow} - \lambda_{\downarrow}$ and dashed arrows denote
bosonic zero-order Green's functions: doublon $g_{20}(\omega_{m})$ or
magnon $g_{\sigma\bar{\sigma}}(\omega_{m})$.

Expression for six-vertex contains the contributions which can be
presented by the following diagrams:
\begin{equation}
  \raisebox{-10pt}[14pt][10pt]{\includegraphics{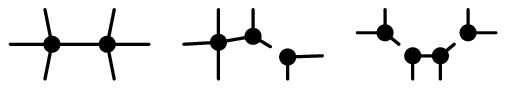}}
  \label{eq26}
\end{equation}
with the internal vertices of the same type as in (\ref{eq25}) and
contribution which can be presented diagrammatically as
\begin{equation}
  \raisebox{-9pt}[14pt][9pt]{\includegraphics{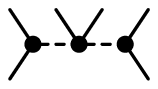}}
  \label{eq27}
\end{equation}
So, we can introduce primitive vertices
\begin{equation}
  \raisebox{-9pt}[14pt][9pt]{\includegraphics{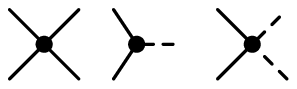}}
  \label{eq28}
\end{equation}
by which one can construct all $n$ vertices in expansion (\ref{eq17})
according to the following rules:
\begin{enumerate}
\item $n$ vertices are constructed by the diagonal Hubbard operator $X^{pp}$
  and zero-order fermionic and bosonic lines connected by primitive vertices
  (\ref{eq28}) specific for each subspace $p$.
\item External lines of $n$ vertices must be of the fermionic type.
\item Diagrams with the loops formed by zero-order fermionic and bosonic
  Green's functions are not allowed because they are already included into
  the formalism, e.g.,
  \(\raisebox{-3pt}[12pt][3pt]{\includegraphics{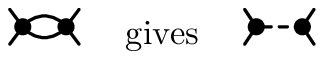}}\).
\end{enumerate}
For $n$ vertices of higher order a new primitive vertices can appear but
we do not check this due to the rapid increase of the algebraic
calculations with the increase of $n$. Diagrams (\ref{eq25}),
(\ref{eq26}), and (\ref{eq27}) topologically are truncated Bethe-lattices
constructed by the primitive vertices (\ref{eq28}) and can be treated as
some generalization of the Hubbard stars \cite{Dongen,Gros1,Gros2} in the
thermodynamical perturbation theory.

It should be noted that each $n$ vertex contains Coulomb interaction $U$
as in primitive vertices (\ref{eq28}) (denoted by dots) as in the
denominators of the zero-order Green's functions (\ref{eq16}). In the
$U\to\infty$ limit, each term in the expressions for $n$ vertices can
diverge but total vertex possesses finite $U\to\infty$ limit when
diagrammatic series of Ref.~\cite{IzyumovLetfulov} are reproduced.

The second problem of calculation of the averages of diagonal $X$
operators is more complicated. One of the ways to solve it is to use
semi-invariant (cumulant) expansions as was done in
Refs.~\cite{IzyumovLetfulov} and \cite{IzyumovPRB} for the $U=\infty$
limit. Another way is to consider the $d=\infty$ limit where new
simplifications appear.

\section{Dynamical mean-field theory}

In general, the grand canonical potential for lattice is connected with
the one for the auxiliary single-site problem of the DMFT by the
expression \cite{BrandtMielsch}
\begin{eqnarray}
  \frac{\Omega}{N} &=& \Omega_{\mathrm{imp}} - \frac1{\beta} \sum_{\nu\sigma} \left\{ \ln
  G_{\sigma}^{(a)} (\omega_{\nu}) \vphantom{\frac{1}{N}} \right. \nonumber\\
&-&\left.   \frac{1}{N} \sum_{\vec{k}} \ln G_{\sigma}
  (\omega_{\nu},\vec{k})\right\}\;.
  \label{eq33}
\end{eqnarray}
On the other hand, we can write for the grand canonical potential for
atomic limit $\Omega_{\mathrm{imp}}$ the same expansion as in (\ref{eq17})
but now we have averages of the products of diagonal $X$ operators at the
same site. According to (\ref{eq12}) we can multiply them and reduce their
product to a single $X$ operator that can be taken outside of the brackets
and exponent in (\ref{eq17}) and its average is equal to
\[
  \langle X^{pp}\rangle_{0} =
  \frac{\E^{-\beta\lambda_{p}}}{\sum_q \E^{-\beta\lambda_{q}}}\;.
\]
Finally,
for the grand canonical potential in atomic limit we get
\begin{equation}
  \Omega_{\mathrm{imp}} = -\frac{1}{\beta} \ln \sum_{p} \E^{-\beta\Omega_{(p)}}\;,
  \label{eq34}
\end{equation}
where
\begin{eqnarray}
  \Omega_{(p)}=
  \lambda_p+\frac1{\beta}\Biggl\{
  \raisebox{-12pt}[15pt][12pt]{\includegraphics[scale=0.7]{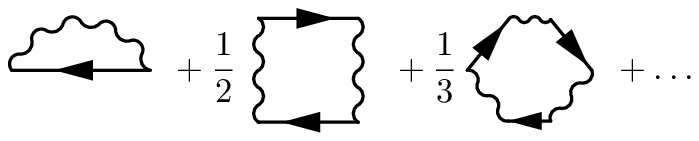}}
  \label{eq35}\\
  \raisebox{-12pt}[14pt][12pt]{\includegraphics[scale=0.7]{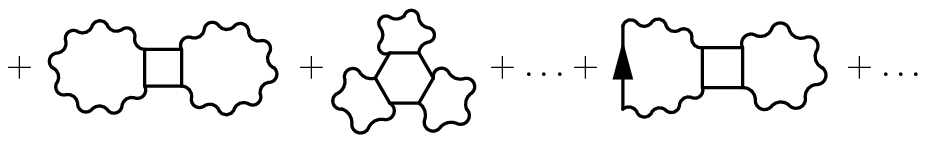}}
  \Biggr\}
  \nonumber
\end{eqnarray}
are the ``grand canonical potentials'' for the subspaces.

Now we can find single-electron Green's function for atomic limit
\begin{eqnarray}
  G_{\sigma}^{(a)} (\tau-\tau') &=& \beta\frac{\delta\Omega_{\mathrm{imp}}}{\delta
  J_{\sigma}(\tau-\tau')} \nonumber\\
&=& \sum_{p} w_{p} G_{\sigma(p)} (\tau-\tau')\;,
  \label{eq36}
\end{eqnarray}
where
\begin{equation}
  G_{\sigma(p)} (\tau-\tau') = \beta\frac{\delta\Omega_{(p)}}{\delta
  J_{\sigma}(\tau-\tau')}
  \label{eq37}
\end{equation}
are single-electron Green's functions for the subspaces characterized by
the ``statistical weights''
\begin{equation}
  w_{p} = \frac{\E^{-\beta\Omega_{(p)}}}{\sum\limits_{q}\E^{-\beta\Omega_{(q)}}}
  \label{eq38}
\end{equation}
and our single-site atomic problem exactly splits into four subspaces
$p=0,2,\downarrow,\uparrow$.

We can introduce irreducible parts of Green's functions in subspaces
$\Xi_{\sigma(p)}(\omega_{\nu})$ by
\begin{equation}
  G_{\sigma(p)} (\omega_{\nu}) = \frac{1}{\Xi_{\sigma(p)}^{-1}(\omega_{\nu}) -
  J_{\sigma}(\omega_{\nu})}\;,
  \label{eq39}
\end{equation}
where
\begin{equation}
  \Xi_{\sigma(p)}(\omega_{\nu})=
  \raisebox{-1.5pt}[21.5pt][1.5pt]{\includegraphics{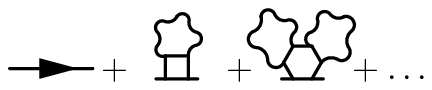}}\;.
\end{equation}

According to the rules of the introduced diagrammatic technique, $n$
vertices are terminated by the fermionic Green's functions [see
(\ref{eq25}), (\ref{eq26}), and (\ref{eq27})] and this allows us to write
a Dyson equation for the irreducible parts and to introduce a self-energy
in subspaces
\begin{equation}
  \Xi_{\sigma(p)}^{-1} (\omega_{\nu}) = g_{\sigma(p)}^{-1} (\omega_{\nu}) -
  \Sigma_{\sigma(p)} (\omega_{\nu})\;,
  \label{eq40}
\end{equation}
where self-energy $\Sigma_{\sigma(p)}(\omega_{\nu})$ depends on the
hopping integral $J_{\sigma'}(\omega_{\nu'})$ only through quantities
\begin{eqnarray}
  \lefteqn{\Psi_{\sigma'(p)} (\omega_{\nu'}) =
  G_{\sigma'(p)} (\omega_{\nu'}) - \Xi _{\sigma'(p)} (\omega_{\nu'})}
\nonumber \\
  &&\quad\equiv
  \Xi^2_{\sigma'(p)} (\omega_{\nu'})J_{\sigma'}(\omega_{\nu'})
\nonumber\\
&& \quad \times  \left\{ 1 +
   \Xi_{\sigma'(p)}(\omega_{\nu'})J_{\sigma'}(\omega_{\nu'})
  + \cdots
  \right\}\;.
  \label{eq41}
\end{eqnarray}
It should be noted, that the total self-energy of the atomic problem is
connected with the total irreducible part by the expression
\begin{equation}
  \Sigma_{\sigma}(\omega_{\nu})=\I\omega_{\nu}+\mu-\Xi_{\sigma}^{-1}(\omega_{\nu})
\end{equation}
and it has no direct connection with the self-energies in the subspaces.

The fermionic zero-order Green's function (\ref{eq23}) can be also
represented in the following form
\begin{equation}
  g_{\sigma(p)}=\frac1{\I\omega_{\nu}+\mu_{\sigma}-Un^{(0)}_{\bar\sigma(p)}}\;,
\end{equation}
where
\begin{equation}
  n^{(0)}_{\sigma(p)}=-\frac{d\lambda_p}{d\mu_{\sigma}}=\left\{\begin{array}{cc}
    0 & \mbox{for } p=0,\bar\sigma \\
    1 & \mbox{for } p=2,\sigma
  \end{array}\right.
\end{equation}
is an occupation of the state $|p\rangle$ by the electron with spin
$\sigma$, and Green's function (\ref{eq39}) can be written as
\begin{eqnarray}
  G_{\sigma(p)}(\omega_{\nu})&=& \Bigl[\I\omega_{\nu}+\mu_{\sigma}-Un^{(0)}_{\bar\sigma(p)}
 \nonumber\\
&-&
\Sigma_{\sigma(p)}(\omega_{\nu})-J_{\sigma}(\omega_{\nu})\Bigr]^{-1}\;.
\label{GFp}
\end{eqnarray}

Now, one can reconstruct expression for the grand canonical potentials
$\Omega_{(p)}$ in subspaces from the known structure of Green's functions.
To do this, we scale hopping integral
\begin{equation}
  J_{\sigma}(\omega_{\nu}) \rightarrow \alpha J_{\sigma}(\omega_{\nu})\;,
  \quad
  \alpha\in [0,1]\;,
\end{equation}
which allows to define the grand canonical potential as
\begin{equation}
  \Omega_{(p)} = \lambda_{p} + \int\limits_{0}^{1} \D\alpha \frac{1}{\beta}
  \sum_{\nu\sigma} J_{\sigma} (\omega_{\nu}) G_{\sigma(p)} (\omega_{\nu}, \alpha)
  \label{eq42}
\end{equation}
and after some transformations one can get
\begin{eqnarray}
  \Omega_{(p)} &=& \lambda_{p} - \frac{1}{\beta} \sum_{\nu\sigma}
  \ln \left[1 - J_{\sigma}(\omega_{\nu}) \Xi_{\sigma(p)}(\omega_{\nu})\right]
\nonumber\\
  &-& \frac{1}{\beta} \sum_{\nu\sigma} \Sigma_{\sigma(p)}(\omega_{\nu})
  \Psi_{\sigma(p)}(\omega_{\nu})
  +\Phi_{(p)}\;,
  \label{eq43}
\end{eqnarray}
where
\begin{equation}\label{Phi}
  \Phi_{(p)}=\frac{1}{\beta}
  \sum_{\nu\sigma}\int\limits_0^1\!\!\D\alpha\, \Sigma_{\sigma(p)}(\omega_{\nu},\alpha)
  \frac{d\Psi_{\sigma(p)}(\omega_{\nu},\alpha)}{d\alpha}
\end{equation}
is some functional, such that its functional derivative with respect to
$\Psi$ produces self-energy:
\begin{equation}
  \beta\frac{\delta\Phi_{(p)}}{\delta\Psi_{\sigma(p)}(\omega_{\nu})}
  =\Sigma_{\sigma(p)}(\omega_{\nu})\;.
\end{equation}
So, if one can find or construct self-energy
$\Sigma_{\sigma(p)}(\omega_{\nu})$ he can find Green's functions and grand
canonical potentials for subspaces and, according to (\ref{eq34}) and
(\ref{eq36}), solve auxiliary single-site problem.

Starting from the grand canonical potential (\ref{eq34}) and (\ref{eq43})
one can get for mean values (\ref{eq10}),
\begin{eqnarray}
  n_{\sigma}&=&\sum_p w_p n_{\sigma(p)}\;,
  \nonumber \\
  n_{\sigma(p)}&=&n^{(0)}_{\sigma(p)}+
  \frac1{\beta}\sum_{\nu}\left[G_{\sigma(p)}(\omega_{\nu})-\Xi_{\sigma(p)}(\omega_{\nu})\right]
\nonumber\\
&-&\frac{\partial\Phi_{(p)}}{\partial\mu_{\sigma}}\;,
  \label{mv}
\end{eqnarray}
where in the last term the partial derivative is taken over the
$\mu_{\sigma}$ not in the fields $\Psi_{\sigma(p)}(\omega_{\nu})$
(\ref{eq41}). The second term in the right-hand side of (\ref{mv}) can be
represented diagrammatically as
\begin{equation}\label{mvPsi}
  \includegraphics{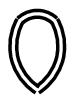}
\end{equation}
and the first contributions into the last term are following
\begin{equation}\label{dPhi}
  \raisebox{-4pt}[28pt][4pt]{\includegraphics{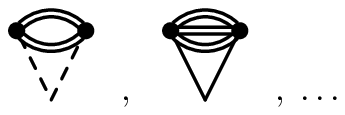}}\;,
\end{equation}
where double lines denote quantities $\Psi_{\sigma(p)}(\omega_{\nu})$.
Loop \raisebox{-5pt}[10pt][5pt]{\includegraphics{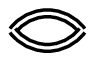}} is
connected with the superconducting or magnon susceptibilities for
subspaces $p=0,2$ or $p=\sigma,\bar\sigma$, respectively.

For the single atom [$J_{\sigma}(\omega_{\nu})=0$] we have $\Phi_{(p)}=0$,
$G_{\sigma(p)}(\omega_{\nu})=\Xi_{\sigma(p)}(\omega_{\nu})=g_{\sigma(p)}(\omega_{\nu})$,
and
\begin{equation}
  n_{\sigma}=\sum_p w_p \frac1{\beta}\sum_{\nu} g_{\sigma(p)}(\omega_{\nu})
  =\sum_p w_p n^{(0)}_{\sigma(p)}\;,
\end{equation}
but in the general case [$J_{\sigma}(\omega_{\nu})\ne0$] we cannot prove
that the sum rule
\begin{equation}
  n_{\sigma}=\frac1{\beta}\sum_{\nu} G^{(a)}_{\sigma}(\omega_{\nu})
\end{equation}
is fulfilled.

\subsection{Falicov--Kimball model}

For the Falicov--Kimball model $J_{\downarrow}(\omega_{\nu})=0$ and
according to (\ref{eq20})
\begin{eqnarray}
  \Sigma_{\uparrow(p)}(\omega_{\nu}) \equiv 0\;; \quad &&
  \Xi_{\uparrow(p)}(\omega_{\nu}) = g_{\uparrow(p)} (\omega_{\nu})\;; \nonumber\\
&&  \Phi_{(p)} \equiv 0
  \label{eq44}
\end{eqnarray}
and
\begin{equation}
  \Omega_{(p)} = \lambda_{p} - \frac{1}{\beta} \sum_{\nu}\ln
  \left[1 - J_{\uparrow}(\omega_{\nu})g_{\uparrow(p)}(\omega_{\nu})\right]\;,
  \label{eq45}
\end{equation}
\begin{eqnarray}
  G_{\uparrow}^{(a)}(\omega_{\nu}) &=& \frac{1-n_{\downarrow}}
  {\I\omega_{\nu}-\lambda_{\uparrow0}-J_{\uparrow}(\omega_{\nu})}
\nonumber\\
&+& \frac{n_{\downarrow}}
  {\I\omega_{\nu}-\lambda_{2\downarrow}-J_{\uparrow}(\omega_{\nu})}\;,
  \label{eq45p}
\end{eqnarray}
\begin{equation}
  n_{\uparrow}=\frac1{\beta}\sum_{\nu} G_{\uparrow}^{(a)}(\omega_{\nu})\;,
  \quad n_{\downarrow}=w_2+w_{\downarrow}\;,
  \label{eq45pp}
\end{equation}
which immediately gives results of \cite{BrandtMielsch} (see also
Ref.~\cite{ShvaikaJPS}). Expression (\ref{eq45p}) is the Matsubara
representation of the Green's function (\ref{eq7.5}) and the second term
in (\ref{eq45}) corresponds to the quantities $Q_{1,2}$ (\ref{eq7.5p}),
which give its real axis representation.

For the Hubbard model there are no exact expression for self-energy but
the set of Eqs.~(\ref{eq39}), (\ref{eq40}), and (\ref{eq43}) allows one to
construct different self-consistent approximations.

\subsection{Alloy-analogy approximation}

The simplest approximation, which can be done, is to put
\begin{equation}
  \Sigma_{\sigma(p)}(\omega_{\nu})=0\;; \quad \Phi_{(p)} = 0
\end{equation}
which gives
\begin{equation}
  \Xi_{\sigma(p)}(\omega_{\nu}) = g_{\sigma(p)} (\omega_{\nu})
\end{equation}
and
\begin{equation}
  \Omega_{(p)} = \lambda_{p} - \frac{1}{\beta} \sum_{\nu\sigma}\ln
  \left[1 - J_{\sigma}(\omega_{\nu})g_{\sigma(p)}(\omega_{\nu})\right]
\end{equation}
and for the Green's function for the atomic problem one can obtain a
two-pole expression
\begin{eqnarray}
  G_{\sigma}^{(a)}(\omega_{\nu}) &=& \frac{w_0+w_{\sigma}}
              {\I\omega_{\nu}-\lambda_{\sigma0}-J_{\sigma}(\omega_{\nu})} \nonumber\\
&+&           \frac{w_2+w_{\bar\sigma}}
              {\I\omega_{\nu}-\lambda_{2\bar\sigma}-J_{\sigma}(\omega_{\nu})}
\label{AAA}
\end{eqnarray}
of the alloy-analogy solution for the Hubbard model, which is a zero-order
approximation within the considered approach. For this approximation, mean
values (\ref{eq10}) are equal to
\begin{eqnarray}
  n_{\sigma} &=& \frac1{\beta} \sum_{\nu} G_{\sigma}^{(a)}(\omega_{\nu}) +
  w_2+w_{\sigma}-\frac{w_0+w_{\sigma}}{\E^{\beta\lambda_{\sigma0}}+\!1}
  - \frac{w_2+w_{\bar\sigma}}{\E^{\beta\lambda_{2\bar\sigma}}+\!1}
  \nonumber\\
  &\neq& \frac1{\beta} \sum_{\nu} G_{\sigma}^{(a)}(\omega_{\nu})
\end{eqnarray}
and, for some values of the chemical potential, they can get unphysical
values: negative or greater then one.

\subsection{Hartree--Fock approximation}

The next possible approximation is to take into account the contribution
from diagram (\ref{mvPsi}) and to construct the equation for the
self-energy in the following form:
\begin{equation}
  \Sigma_{\sigma(p)}(\omega_{\nu})=\frac1{\beta}\sum_{\nu'}
  U\Psi_{\bar\sigma(p)}(\omega_{\nu'})\;,
\end{equation}
which, together with the expression for mean values
\begin{eqnarray}
  n_{\sigma(p)} &=& n^{(0)}_{\sigma(p)} + \frac{1}{\beta} \sum\limits_{\nu}
  \Psi_{\sigma(p)}(\omega_{\nu})
 \nonumber  \\
  &=& n^{(0)}_{\sigma(p)}-\frac12+
  \frac12\tanh\frac{\beta}2 \left[Un_{\bar\sigma(p)}-\mu_{\sigma}\right]
 \nonumber \\
&+& \frac1{\beta}\sum_{\nu'}
  G_{\sigma(p)}(\omega_{\nu'})\;,
\label{mvHF}
\end{eqnarray}
gives for the Green's function in the subspaces expression in the
Hartree--Fock approximation:
\begin{equation}\label{GFHF}
  G_{\sigma(p)}(\omega_{\nu})=\frac1{\I\omega_{\nu}+\mu_{\sigma}-Un_{\bar\sigma(p)}
  -J_{\sigma}(\omega_{\nu})}\;.
\end{equation}
Now, grand canonical potentials in the subspaces are equal
\begin{eqnarray}
   \Omega_{(p)}&=&\lambda_p-\frac1{\beta}\sum_{\nu\sigma}
   \ln\left[1-J_{\sigma}(\omega_{\nu})\Xi_{\sigma(p)}(\omega_{\nu})\right]
   \nonumber\\
   &&-U \left(n_{\sigma(p)}-n_{\sigma(p)}^{(0)}\right)
   \left(n_{\bar\sigma(p)}-n_{\bar\sigma(p)}^{(0)}\right)\;,
   \label{Phi_H-F}\\
   \Phi_{(p)}&=&U \left(n_{\sigma(p)}-n_{\sigma(p)}^{(0)}\right)
   \left(n_{\bar\sigma(p)}-n_{\bar\sigma(p)}^{(0)}\right)
   \nonumber
\end{eqnarray}
and for the single-site Green's function (\ref{eq36}) one can obtain a
four-pole structure
\begin{equation}\label{GFHFt}
  G^{(a)}_{\sigma}(\omega_{\nu})=\sum_p
  \frac{w_p}{\I\omega_{\nu}+\mu_{\sigma}-Un_{\bar\sigma(p)}
  -J_{\sigma}(\omega_{\nu})}\;.
\end{equation}
Expression (\ref{GFHFt}), in contrast to the alloy-analogy solution
(\ref{AAA}), possesses the correct Hartree--Fock limit for small Coulomb
interaction $U\ll t$:
\begin{equation}
  G^{(a)}_{\sigma}(\omega_{\nu})=
  \frac1{\I\omega_{\nu}+\mu_{\sigma}-Un_{\bar\sigma} - J_{\sigma}(\omega_{\nu})}\;,
\end{equation}
when $w_p\approx 1/4$ and $n_{\sigma(p)}\approx n_{\sigma}=
\Theta\sum_{\nu} G^{(a)}_{\sigma}(\omega_{\nu})$. On the other hand, in
the same way as an alloy-analogy solution, it describes the
metal--insulator transition with the change of $U$.

\begin{figure}[htbp]
\noindent\null\hfill\includegraphics[width=0.4\textwidth]{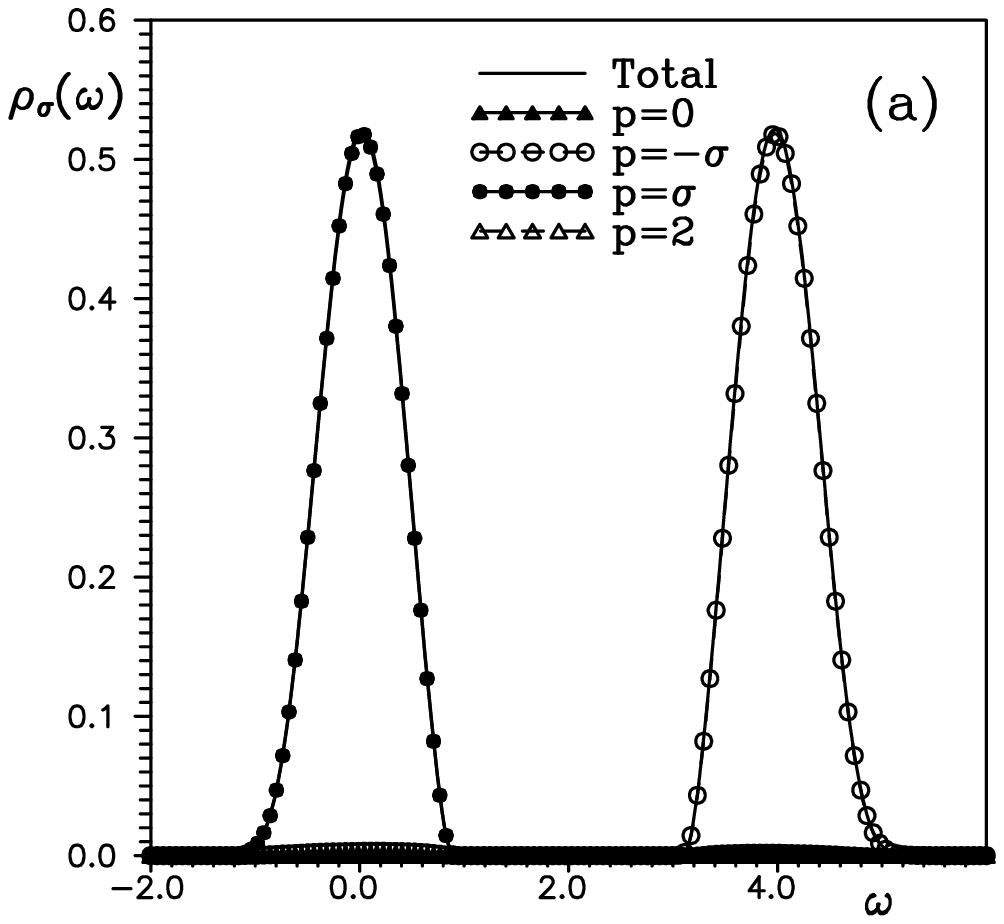}\hfill\null
\null\hfill\includegraphics[width=0.4\textwidth]{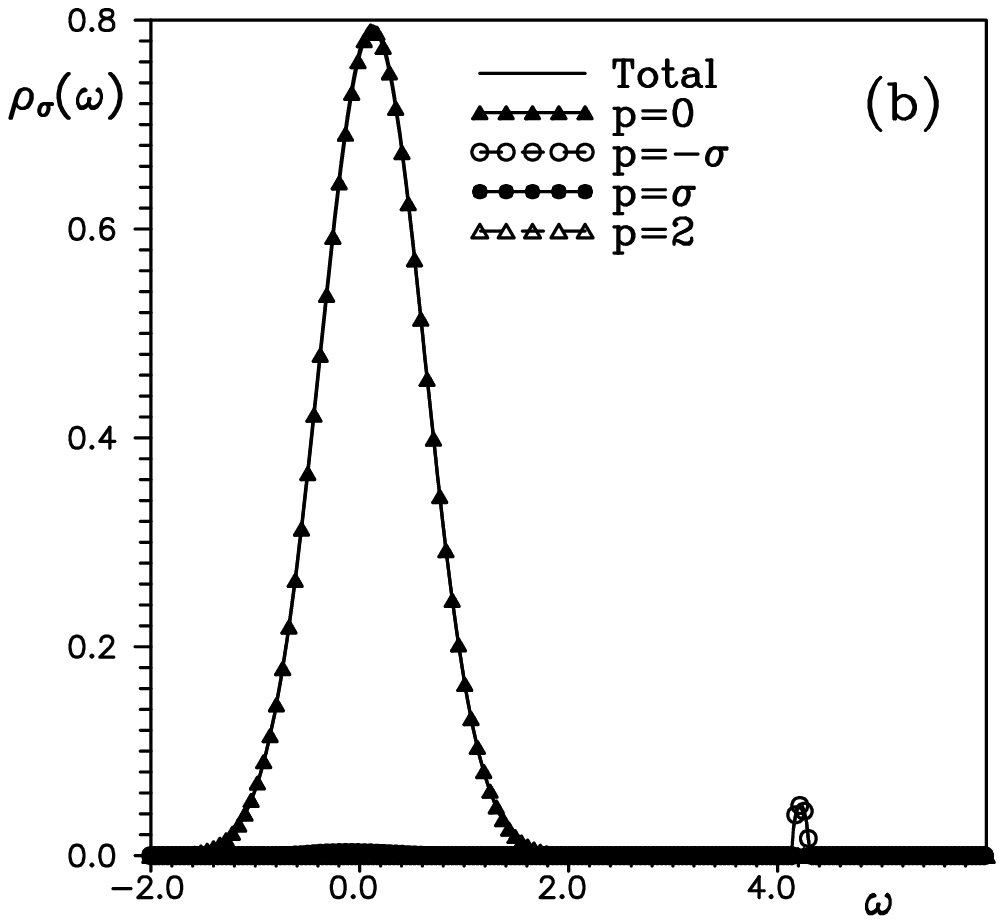}\hfill\null\\
[1em]
\null\hfill\includegraphics[width=0.4\textwidth]{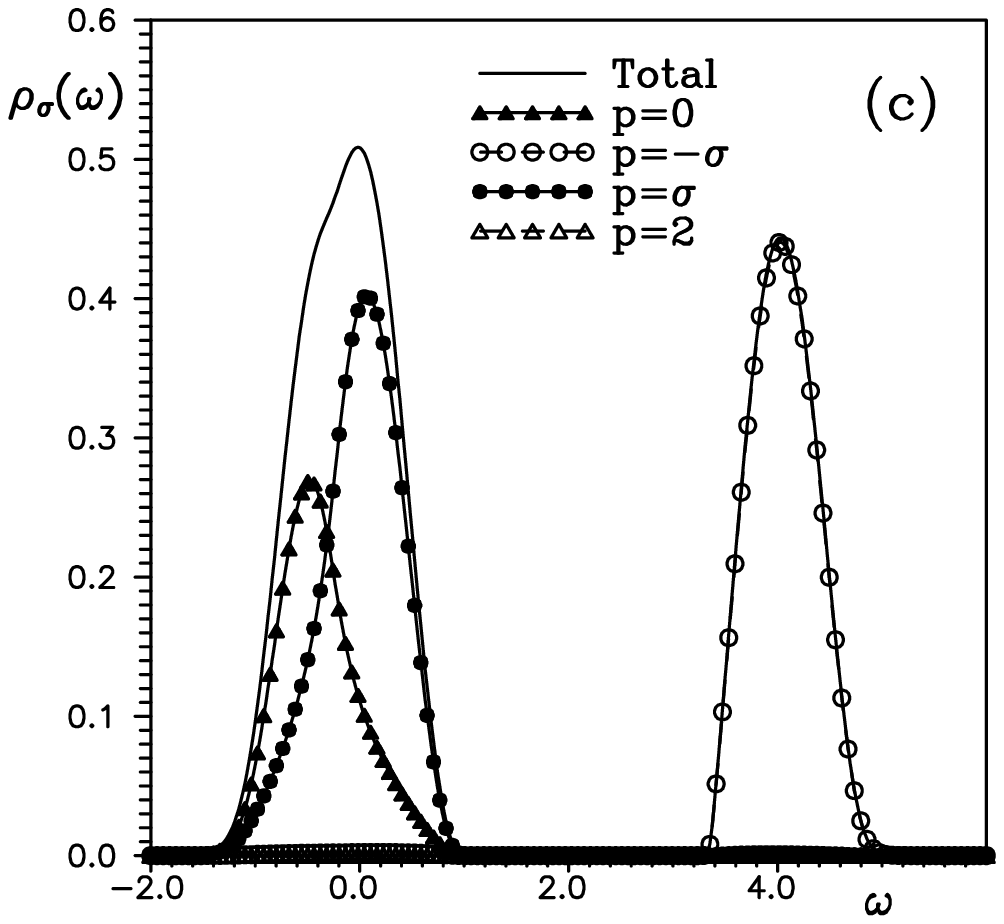}\hfill\null
\null\hfill\includegraphics[width=0.4\textwidth]{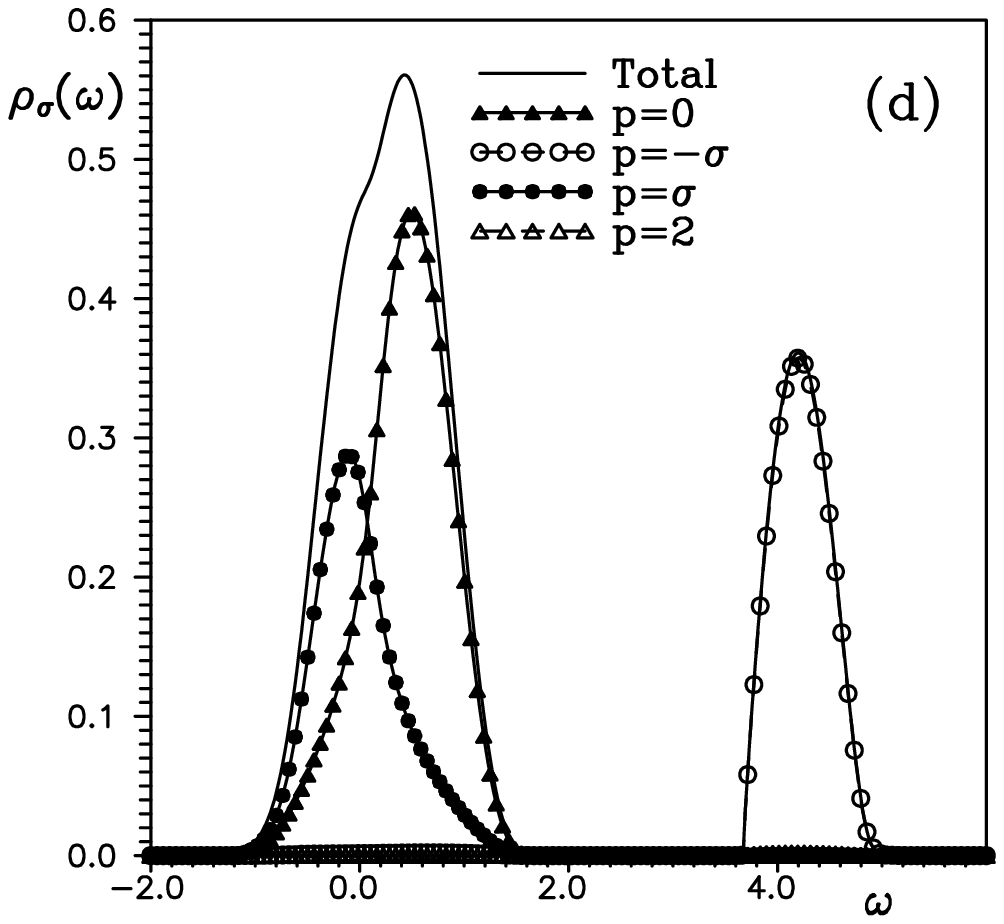}\hfill\null
\bigskip
\caption{Spectral weight function $\rho_{\sigma}(\omega)$
(\protect\ref{SWFt}): total and for each subspace, for the different
chemical potential values: (a) $\mu= U/2$, $n=1$; (b) $\mu=-1$, $n=0.07$;
(c) $\mu=0.01$, $n=0.72$; (d) $\mu=-0.01$, $n=0.66$ ($U=4$, $T=0.2$)}
\label{ImG_f}
\end{figure}

\begin{figure}[htbp]
\noindent\null\hfill\includegraphics[width=0.4\textwidth]{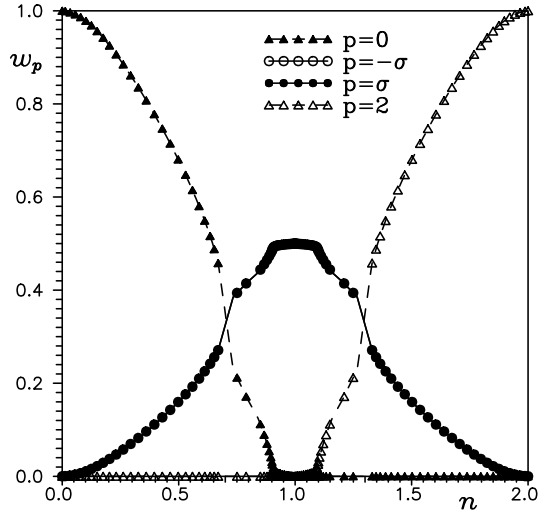}\hfill\null
\bigskip
\caption{Statistical weights of the subspaces $w_p$ (\protect\ref{eq38})
as functions of the electron concentration ($U=4$, $T=0.2$)} \label{wp_n}
\end{figure}

In Fig.~\ref{ImG_f} the frequency distribution of the total spectral
weight function
\begin{equation}\label{SWFt}
  \rho_{\sigma}(\omega)=\frac1{\pi}\Im G^{(a)}_{\sigma}(\omega-\I 0^+)
\end{equation}
as well as contributions into it from the subspaces [separate terms in
(\ref{GFHFt})] are presented for the different electron concentration
(chemical potential) values. One can see, that the spectral weight
function contains two peaks, which correspond to the two Hubbard bands.
Each band is formed by the two close peaks: $p=0$ and $\sigma$ for the
lower Hubbard band and $p=2$ and $\bar\sigma$ for the upper one, with
weights $w_p$ (\ref{eq38}). The main contributions come (see
Fig.~\ref{wp_n}) from the subspaces $p=0$ for the low electron
concentrations ($n<2/3$, $\mu<0$), $p=2$ for the low hole concentrations
($2-n<2/3$, $\mu>U$) and $p=\sigma,\bar\sigma$ for the intermediate
values. For the small electron or hole concentrations, the Green's
function for the atomic problem (\ref{GFHFt}) possesses correct
Hartree--Fock limits too.

%%Figure 1

%%Figure 2

Such four-pole structure of the single-electron Green's function can be
obtained also for the one-dimensional chain with the $N=2$ periodic
boundary condition (see Appendix in Ref.~\cite{ShvaikaPRB}), which is
equivalent to the two-site problem considered by Harris and Lange
\cite{H_L}. Here, two poles correspond to the noninteracting electrons or
holes, which hope over the empty sites, and give the main contribution for
small concentrations. The other two poles give the main contribution close
to half-filling and correspond to the hopping of the strongly-correlated
electrons over the resonating valence bond (RVB) states.

So, one can suppose that the Hubbard model describes strongly-correlated
electronic systems that contain four components (subspaces). Subspaces
$p=0$ and $p=2$ describe the Fermi-liquid component (electron and hole,
respectively) which is dominant for the small electron and hole
concentrations, when the chemical potential is close to the bottom of the
lower band and top of the upper one. On the other hand, subspaces
$p=\uparrow$ and $\downarrow$ describe the non-Fermi-liquid (strongly
correlated, e.g., RVB) component, which is dominant close to half-filling.
The plateau at half filling for $w_p$ ($p=\uparrow,\downarrow$) can be
associated with the antiferromagnetic phase. Within the considered
Hartree--Fock approximation, at $n\approx 2/3$ and $2-n\approx 2/3$, we
have transition between these two regimes: Fermi liquid and non-Fermi
liquid. It reminds us the known properties of the high-$T_c$ compounds,
where for the nondoped case ($n=1$) compounds are in the antiferromagnetic
dielectric state, then for small doping the non-Fermi-liquid behavior is
observed (underdoped case $n\lesssim1$) and after some optimal doping
value, the properties of the compound sharply change from the non-Fermi to
the Fermi liquid (overdoped case).

The results presented in Figs.~\ref{ImG_f} and \ref{wp_n} are obtained for
relatively high temperature. With the temperature decrease, on the one
hand, the transition between the Fermi and non-Fermi liquid becomes sharp
and, on the other hand, for some chemical potential values there can be
three solutions of (\ref{mvHF}) with two of them corresponding to the
phase-separated states. The consideration of the phase separation in the
Hubbard model is not a topic of this paper and will be the subject of
further investigations.

\begin{figure}[htbp]
 \centerline{\includegraphics[width=.47\textwidth]{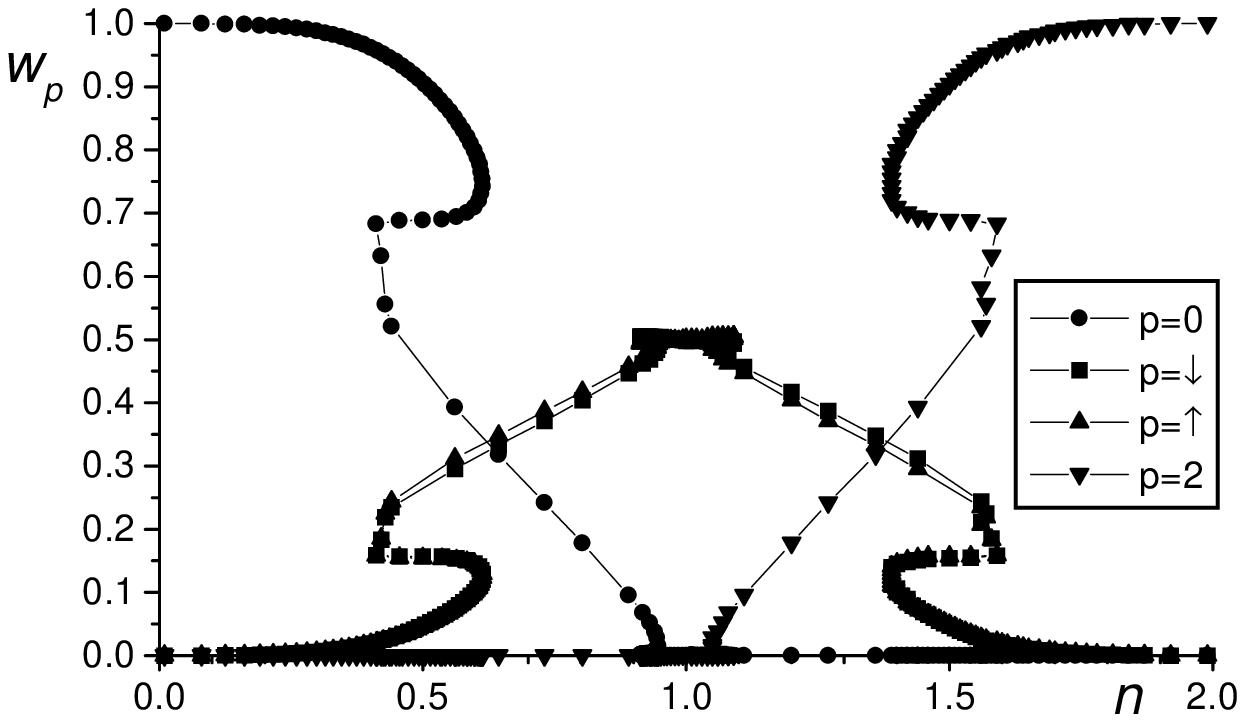}}
 \bigskip
 \centerline{\includegraphics[width=.47\textwidth]{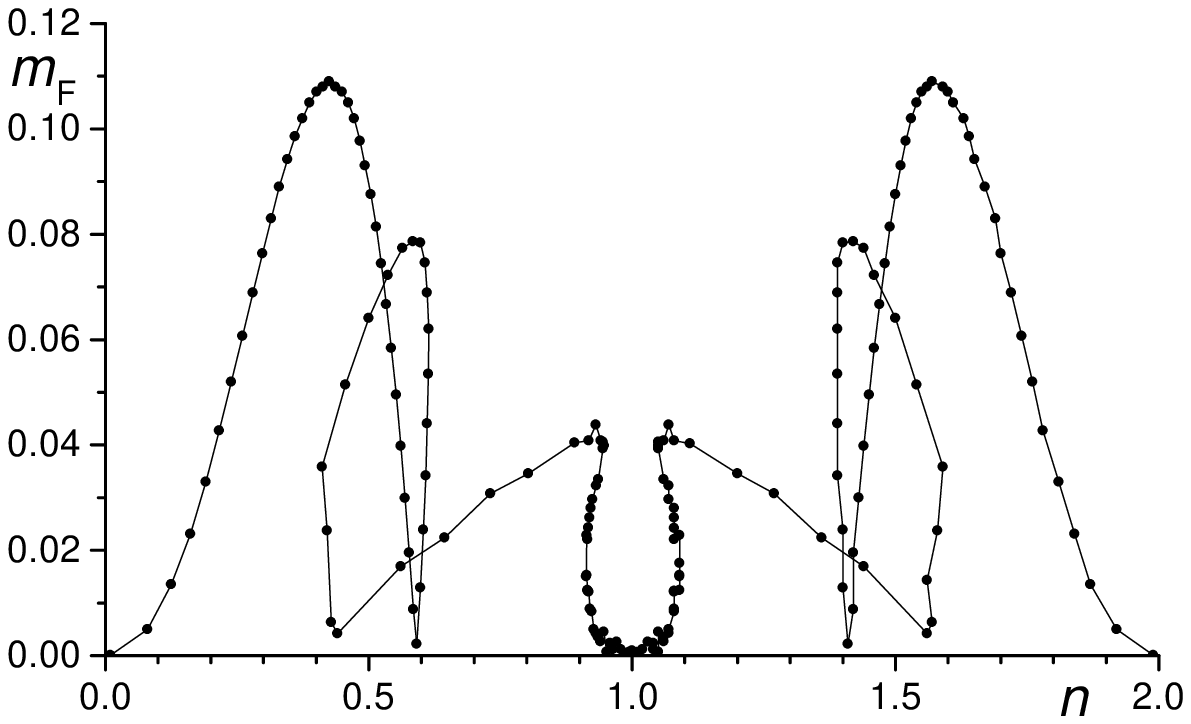}\quad
 \includegraphics[width=.47\textwidth]{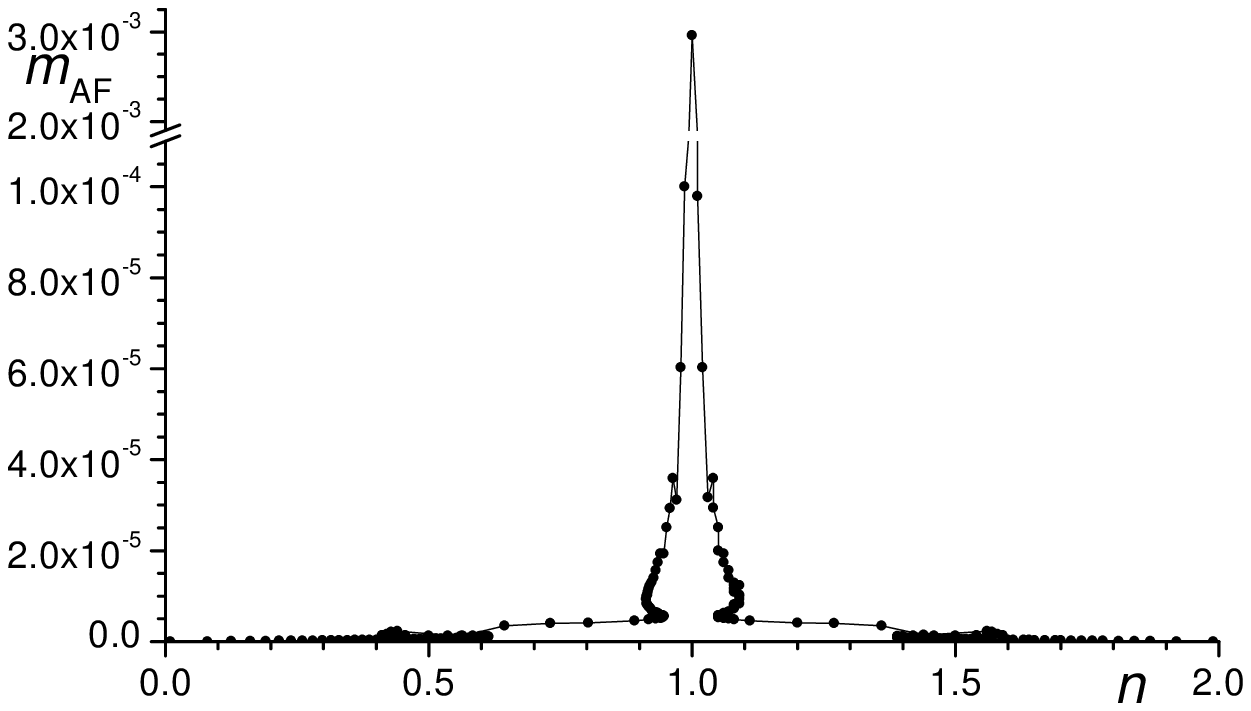}}
 \caption{Statistical weights of subspaces $w_p$
 and ferromagnetic $m_{F}$ and antiferromagnetic $m_{AF}$ order
 parameters vs electron concentration for $U=1.56$, $T=0.14$}
 \label{FerrOP}
\end{figure}

At low temperatures, besides the plateau on the concentration dependence
of $w_p$ for $p=\sigma,\bar\sigma$ at half filling, also the plateau for
the statistical weights of subspaces $p=0,2$ are developed at low electron
and hole concentrations, see Fig.~\ref{FerrOP}. The $p=0$ and $p=2$
components for the low electron or hole concentrations are in the
ferromagnetic state, while the non-Fermi-liquid one is antiferromagnetic
(AF) close to half-filling \cite{ShvaikaCMP}. For the intermediate
concentration values the picture is very complicated, even frustrated. It
is due to the fact that equations for the mean values (\ref{mvHF}) have
several solutions in this region, which, on the other hand, are mutually
connected with the dynamical mean field $J_{\sigma}(\omega_{\nu})$. It is
difficult to determine the ground state for this, possibly ``pseudo-gap'',
region, which is located between the ferromagnetic and antiferromagnetic
phases.

%%Figure 3

\begin{figure}[htbp]
\centerline{\includegraphics[width=.7\textwidth]{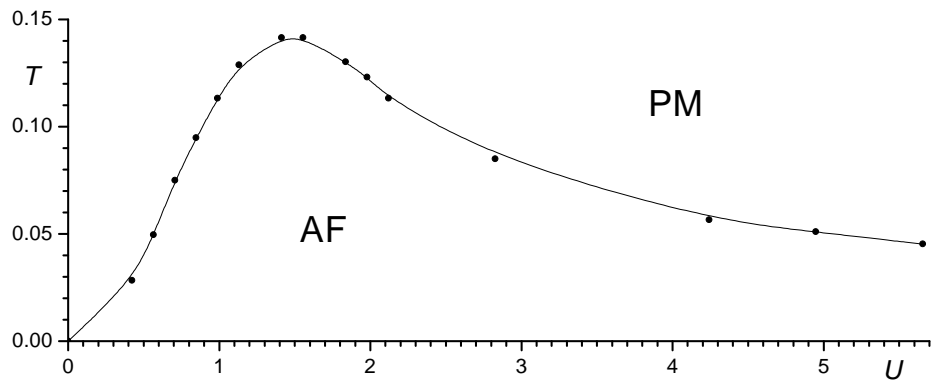}}
 \caption{Phase diagram $(T,U)$ at half-filling $n=1$ (AF~--
 antiferromagnetic phase, PM~-- paramagnetic phase)}
 \label{T-U}
\end{figure}

In Fig.~\ref{T-U} we presented the phase diagram $(T,U)$~-- the
temperature of the AF ordering vs correlation energy $U$, which is in a
qualitative agreement with the results of Refs.~\cite{Pruschke,T-U1,T-U2}
and reproduces the results of the Hartree--Fock theory and mean field
approximation for $U\ll t$ and $U\gg t$, respectively. Our results for the
AF critical temperature for small $U$ are higher then the one of the
Quantum Monte Carlo simulations \cite{Pruschke} by about a factor at three
that describes the reduction of the Hartree--Fock solution by the lowest
order quantum fluctuations \cite{Dongen}.

%%Figure 4

\subsection{Beyond the Hartree--Fock approximation}

Self-energy in the Hartree--Fock approximation [see Eq.~(\ref{GFHF})]
describes some self-consistent shift of the initial energy levels and does
not depend on the frequency. All other improvements of the expression for
self-energy add the frequency dependent contributions. To see this, let us
consider the contribution into the mean values from the first diagram in
(\ref{dPhi}). This diagram originates from the following skeletal diagram
\begin{equation}\label{skelet}
  \raisebox{-13pt}[17pt][13pt]{\includegraphics{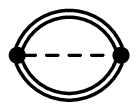}}
\end{equation}
in the diagrammatic expansion for functional $\Phi_{(p)}$. On the other
hand, such a skeletal diagram produces additional contribution into the
self-energy
\begin{equation}\label{SigmaF}
  \raisebox{-4pt}[17pt][4pt]{\includegraphics{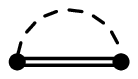}}
\end{equation}
which is frequency dependent. Also, in order to get a self-consistent set
of equations, we introduce renormalized bosonic Green's functions
\[
  D_{20(p)}(\omega_m)= \frac1{\I\omega_m-\tilde\lambda_{20(p)}}\;;
\]
\[
  D_{\sigma\bar\sigma(p)}(\omega_m)=
  \frac1{\I\omega_m-\tilde\lambda_{\sigma\bar\sigma(p)}}\;,
\]
\begin{eqnarray}
  \label{D02}
  \tilde\lambda_{20(p)}&=&\lambda_{20}+U\frac1{\beta}\sum_{\nu\sigma}
  \Psi_{\sigma(p)}(\omega_{\nu})\;,
  \\ \nonumber
  \tilde\lambda_{\sigma\bar\sigma(p)}&=&\lambda_{\sigma\bar\sigma}+
  U\frac1{\beta}\sum_{\nu}\left(\Psi_{\bar\sigma(p)}(\omega_{\nu})
              -\Psi_{\sigma(p)}(\omega_{\nu})\right)\;.
\end{eqnarray}

Finally, for the Green's function (\ref{GFp}) we get the general
representation
\begin{eqnarray}
G_{\sigma(p)}(\omega_{\nu})&=& \Bigl[
\I\omega_{\nu}+\mu_{\sigma}-Un_{\bar\sigma(p)} \nonumber\\
&-&
\widetilde\Sigma_{\sigma(p)}(\omega_{\nu})-J_{\sigma}(\omega_{\nu})\Bigr]^{-1}\;,
\label{GFpf}
\end{eqnarray}
where the Hartree--Fock contribution $Un_{\bar\sigma(p)}$ is extracted and
$\widetilde\Sigma_{\sigma(p)}(\omega_{\nu})$ is a frequency dependent part
of the self-energy, which within the considered approximation is equal
\begin{eqnarray}
&&  \widetilde\Sigma_{\sigma(p)}(\omega_{\nu})= \nonumber\\
&& \quad
\pm\frac{U^2}{\beta}\sum_{\nu'}\tilde{D}_{\sigma\bar\sigma(p)}(\omega_{\nu},\omega_{\nu'})
  \Psi_{\bar\sigma(p)}(\omega_{\nu'})\;,
\label{SE02}
\end{eqnarray}
where
\begin{equation}
  \tilde{D}_{\sigma\bar\sigma(p)}(\omega_{\nu},\omega_{\nu'})=\left\{\begin{array}{cc}
    D_{20(p)}(\omega_{\nu+\nu'}) & \mbox{ for } p=0,2 \\
    D_{\sigma\bar\sigma(p)}(\omega_{\nu-\nu'}) & \mbox{ for }
    p=\sigma,\bar\sigma
  \end{array}\right.\;.
\end{equation}
Now, mean values (\ref{mv}) are equal
\begin{eqnarray}
 n_{\sigma(p)} &=& n^{(0)}_{\sigma(p)} + \frac{1}{\beta}
\sum\limits_{\nu}
  \Psi_{\sigma(p)}(\omega_{\nu})
\nonumber \\
&\pm& \frac{1}{\beta^{2}} \sum\limits_{\nu\nu'}
  U^2 \tilde{D}^2_{\sigma\bar\sigma(p)}(\omega_{\nu},\omega_{\nu'})
\nonumber\\
&\times& \Psi_{\sigma(p)} (\omega_{\nu})
  \Psi_{\bar\sigma(p)} (\omega_{\nu'})
\label{eq49}
\end{eqnarray}
and for the functional (\ref{Phi}) in the grand canonical potentials in
the subspaces we obtain the following expression
\begin{eqnarray}\label{Omega02}\label{OmegaUD}
  \Phi_{(p)}&=&\frac1{\beta^2}\sum_{\nu\nu'}
  \left[U\pm U^2\tilde{D}_{\sigma\bar\sigma(p)}(\omega_{\nu},\omega_{\nu'})\right]
\nonumber\\
&\times&  \Psi_{\sigma(p)}(\omega_{\nu})
\Psi_{\bar\sigma(p)}(\omega_{\nu'})\;.
\end{eqnarray}
Expression (\ref{OmegaUD}) besides the Hartree--Fock contribution
(\ref{Phi_H-F}) contains also the contribution from the skeletal diagram
(\ref{skelet}).

In order to analyze the structure of the poles in (\ref{GFpf}), an
analytical continuation of the expression for
$\widetilde\Sigma_{\sigma(p)}(\omega_{\nu})$ from the imaginary axis to
the real one should be done. To do it, we use the well-known identity
\begin{equation}
  \frac1{\beta}\sum_{\nu}\frac{\E^{\I\omega_{\nu} 0^+}}{\I\omega_{\nu}-\lambda}=
  \pm n_\pm(\lambda)\;,
\end{equation}
which follows from (\ref{eq15}), and analytical properties of the Green's
function
\begin{equation}
  G_\sigma(z)=\frac1{\pi}\int\limits_{-\infty}^{+\infty}\!\! \D\omega
  \frac{\Im G_\sigma(\omega-\I 0^+)}{z-\omega}\;.
\end{equation}
Green's functions in the subspaces $G_{\sigma(p)}(z)$, irreducible parts
$\Xi_{\sigma(p)}(z)$, and dynamical mean-field $J_{\sigma}(z)$ all possess
the same analytical properties. Finally, we get the following expressions:
\begin{eqnarray}
  \tilde{\Sigma}_{\sigma(p)}(z)&=&\pm \frac{U^2}{\pi}
  \int\limits_{-\infty}^{+\infty}\!\!
  \D\omega\;
  n_+(\omega)\;\frac{\Im\Psi_{\bar\sigma(p)}(\omega-\I 0^+)}{z+\omega-\tilde\lambda_{20(p)}}
\nonumber \\
  && \mp n_-(\tilde\lambda_{20(p)})\;U^2
  \Psi_{\bar\sigma(p)}(\tilde\lambda_{20(p)}-z)
\label{Sigma02}
\end{eqnarray}
for subspaces $p=0,2$ and
\begin{eqnarray}
  \tilde{\Sigma}_{\sigma(p)}(z)&=&\pm \frac{U^2}{\pi}
  \int\limits_{-\infty}^{+\infty}\!\! \D\omega\;
  n_+(\omega)\;\frac{\Im\Psi_{\bar\sigma(p)}(\omega-\I0^+)}
  {z-\omega-\tilde\lambda_{\sigma\bar\sigma(p)}}
  \label{SigmaUD} \\
  && \mp\left[n_-(\tilde\lambda_{\sigma\bar\sigma(p)})+1\right]\,U^2
  \Psi_{\bar\sigma(p)}(z-\tilde\lambda_{\sigma\bar\sigma(p)})
  \nonumber
\end{eqnarray}
for $p=\sigma,\bar\sigma$. Analytical continuation of expressions
(\ref{eq49}) and (\ref{OmegaUD}) can be done in the same way. One can see,
that contributions (\ref{Sigma02}) and (\ref{SigmaUD}) diverge in the
paramagnetic phase close to half filling when $\tilde\lambda_{20(p)}=0$
and $\tilde\lambda_{\sigma\bar\sigma(p)}=0$, respectively, which is an
unphysical result.

So, we cannot include into the consideration only one contribution from
diagram (\ref{skelet}) but one have to consider, besides the fermionic
loops, also the bosonic ones \cite{ShvaikaAPPB} which correspond to the
creation and annihilation of the doublons (pairs of electrons), described
by the $\hat{X}^{20}$ and $\hat{X}^{02}$ operators, for subspaces $p=0,2$
and magnons, described by the $\hat{X}^{\uparrow\downarrow}$ and
$\hat{X}^{\downarrow\uparrow}$ operators, for $p=\uparrow,\downarrow$. The
such loop contributions of bosonic excitations can be summed up and one
can obtain
\begin{eqnarray}\label{appb1}
  \Phi_{(p)}&=&\frac1{\beta}\sum_m \ln
\bigl[1-U \nonumber\\
&\times& \left(1\pm
U\tilde{D}_{\sigma\bar\sigma(p)}(\omega_m)\right)
  \tilde\chi_{\sigma\bar\sigma(p)}(\omega_m)\bigr] \;,
\end{eqnarray}
where
\begin{eqnarray}\label{appb2}\label{appb3}
  \tilde{D}_{\sigma\bar\sigma(p)}(\omega_m) &=& D_{20(p)}(\omega_m)\;,
\\
  \tilde\chi_{\sigma\bar\sigma(p)}(\omega_m) &=&-\frac1{\beta}\sum_{\nu}
  \Psi_{\sigma(p)}(\omega_{\nu}) \Psi_{\bar\sigma(p)}(\omega_{m-\nu})
\nonumber
\end{eqnarray}
for subspaces $p=0,2$ and
\begin{eqnarray}\label{appb4}\label{appb5}
  \tilde{D}_{\sigma\bar\sigma(p)}(\omega_m) &=& D_{\sigma\bar\sigma(p)}(\omega_m)\;,
\\
  \tilde\chi_{\sigma\bar\sigma(p)}(\omega_m) &=& -\frac1{\beta}\sum_{\nu}
  \Psi_{\sigma(p)}(\omega_{\nu}) \Psi_{\bar\sigma(p)}(\omega_{\nu-m})
\nonumber
\end{eqnarray}
for subspaces $p=\sigma,\bar\sigma$. Expression (\ref{OmegaUD}) is the
first term of the expansion of functional (\ref{appb1}) in the series over
$\tilde\chi_{\sigma\bar\sigma(p)}(\omega_m)$.

Now we obtain for mean values the following expression
\begin{eqnarray}\label{appb6}
  n_{\sigma(p)} &=& n_{\sigma(p)}^{(0)}+
  \frac1{\beta}\sum_{\nu}\Psi_{\sigma(p)}(\omega_{\nu})
\\ \nonumber
&\mp&
\frac1{\beta}\sum_m\frac{U^2\tilde{D}_{\sigma\bar\sigma(p)}^2(\omega_m)
  \tilde\chi_{\sigma\bar\sigma(p)}(\omega_m)}
  {1-U\left(1\pm U\tilde{D}_{\sigma\bar\sigma(p)}(\omega_m)\right)
  \tilde\chi_{\sigma\bar\sigma(p)}(\omega_m)}
\end{eqnarray}
and self-energy contains the frequency dependent part
\begin{equation}\label{appb7}
  \Sigma_{\sigma(p)}(\omega_{\nu})=
  U\left(n_{\bar\sigma(p)}-n_{\bar\sigma(p)}^{(0)}\right)+
  \tilde\Sigma_{\sigma(p)}(\omega_{\nu})\;,
\end{equation}
\begin{eqnarray}\nonumber
  &&\tilde\Sigma_{\sigma(p)}(\omega_{\nu})=U^2
  \frac1{\beta}\sum_m
  \\
  &&\frac{\left(1\pm U\tilde{D}_{\sigma\bar\sigma(p)}(\omega_m)\right)
  \tilde\chi_{\sigma\bar\sigma(p)}(\omega_m)\pm\tilde{D}_{\sigma\bar\sigma(p)}(\omega_m)}
  {1-U\left(1\pm U\tilde{D}_{\sigma\bar\sigma(p)}(\omega_m)\right)
  \tilde\chi_{\sigma\bar\sigma(p)}(\omega_m)}
  \nonumber\\ \label{appb8}
  &&\times\left\{
  \begin{array}{rl}
  \Psi_{\bar\sigma(p)}(\omega_{m-\nu}), & \mbox{ for } p=0,2 \\
  \Psi_{\bar\sigma(p)}(\omega_{\nu-m}), & \mbox{ for } p=\sigma,\bar\sigma
  \end{array}
  \right. \;,
\end{eqnarray}
that describes the contributions from the doublons (charge fluctuations)
for the Fermi liquid component ($p=0,2$) and magnons (spin fluctuations)
for the non-Fermi liquid one ($p=\uparrow,\downarrow$) with the
renormalized spectrum determined by the zeros of denominator in
(\ref{appb8}).

Expression (\ref{appb1}) for functional $\Phi_{(p)}$ has the same form as
the correction to free energy in the theory of the self-consistent
renormalization (SCR) of spin fluctuations by Moriya \cite{MoriyaBook}.
But in our case it describes contributions from the single-site bosonic
(spin or charge) fluctuations with specific renormalization functions
different for different subspaces. Spin fluctuations give the main
contribution close to half filling in the non-Fermi liquid regime but for
small electron ($n\ll1$) or hole ($2-n\ll1$) concentrations the
contributions from the charge fluctuations must be taken into account.

\section{Concluding remarks}

An analytical approaches for the solution of the effective single site
problem in the DMFT method for the Hubbard-type models described in this
article are based on the strong coupling scheme that considers the strong
local interaction as reference system. For the first one, it corresponds
to the selection of the Hubbard operators as basis for the projection
procedure for Green's functions while in the second one the perturbation
theory over electron hopping is used. Both of them have their advantages.

The equation of motion method together with the averaging over the
auxiliary Fermi field gives an approximate interpolating scheme that in
specific cases includes a number of known approximations for the Hubbard
and similar models. An examples where the proposed approach gives exact
results are given (Falicov--Kimball and simplified pseudospin-electron
models).

At the same time, the applied procedure of the irreducible Green's
functions introduction and different time decoupling appears to be too
simple to obtain the 4-pole structure for the single-site electron Green's
function. An inclusion only of the Fermi-type single-site Hubbard
operators in the basis at the formulation of the equations of motion
produces the 2-pole Green's function and only the extension of the basis
and application of the projection and decoupling procedures to the higher
order functions probably can be able to reveal the more complicated
structure of function $G_{\sigma}^{(a)}(\omega)$. Besides, this way
requires the consideration of the retarded effective interactions formed
by the auxiliary $\xi$-field.

Nevertheless, it should be mentioned, that the simplicity and
accessibility of such approach based on the equation of motion method
makes it attractive for the approximate analytical considerations. It
seems useful to apply it to the problems which have been considered up to
now by means of numerical methods (or can be solved exactly only
numerically). It should be noted, that the calculation of the electron
mean occupation values (and derivation of the equation for the chemical
potential), as well as the determination of the grand canonical potential
within the equation of motion scheme for the Green's functions are
elucidated only partially in this work. It will be the subject of a
separate publication.

The second approach considered in this article uses for the Hubbard-type
models a finite-temperature perturbation theory scheme in terms of
electron hopping, which is based on the Wick's theorem for Hubbard
operators and is valid for arbitrary values of $U$ ($U<\infty$).
Diagrammatic series contain single-site vertices, which are irreducible
many-particle Green's functions for unperturbated single-site Hamiltonian,
connected by hopping lines. Applying the Wick's theorem for Hubbard
operators has allowed us to calculate these vertices and it is shown that
for each vertex the problem splits into subspaces with ``vacuum states"
determined by the diagonal (projection) operators and only excitations
around these ``vacuum states" are allowed. The vertices possess a finite
$U\to\infty$ limit when diagrammatic series of the strong-coupling
approach \cite{IzyumovLetfulov,IzyumovPRB} are reproduced. The rules to
construct diagrams by the primitive vertices are proposed.

In the limit of infinite spatial dimensions the total auxiliary
single-site problem exactly (naturally) splits into subspaces (four for
Hubbard model) and a considered analytical scheme allows to build a
self-consistent Baym--Kadanoff-type theory for the Hubbard model. Some
analytical results are given for simple approximations: an alloy-analogy
approximation, when two-pole structure for Green's function is obtained,
which is exact for the Falicov--Kimball model, and the Hartree--Fock-type
approximation, which results in the four-pole structure for the Green's
function. Expanding beyond the Hartree--Fock approximation calls for the
considering of the frequency dependent contributions into the self-energy
connected with the self-consistently renormalized spin and charge
fluctuations.

In general, the expression
\begin{equation}
  G^{(a)}_{\sigma}(\omega_{\nu}) =\!\sum_p
  \frac{w_p}{\I\omega_{\nu}+\mu_{\sigma}-Un_{\bar\sigma(p)}
  -\widetilde\Sigma_{\sigma(p)}(\omega_{\nu})-J_{\sigma}(\omega_{\nu})}
\end{equation}
gives an exact four-pole structure for the single-site Green's function of
the effective atomic problem. In (\ref{eq40}) zero-order Green's functions
(\ref{eq23}) are the same for the subspaces $p=0,\sigma$ and
$p=2,\bar\sigma$, respectively, and correspond to the two-pole solution of
the one-site problem without hopping. Switching on of the electron hopping
splits these two poles and the value of splitting is determined by the
values of the self-energy parts in the subspaces, which describe the
contributions from the different scattering processes. Alloy-analogy
approximation neglects such scattering processes
($\Sigma_{\sigma(p)}(\omega_{\nu})=0$) which results in the two-pole
structure for the Green's functions (\ref{AAA}). But, in general, Green's
functions possess four-pole structure and even the Hartree--Fock
approximation (\ref{GFHFt}) clearly shows it.

It should be noted that the four-pole structure of the Green's function
for the atomic problem might not result in the four bands of the spectral
weight function (see Fig.~\ref{ImG_f}). The presented consideration allows
us to suppose that each pole describes contributions from the different
components (subspaces) of the electronic system: Fermi liquid (subspaces
$p=0,2$) and non-Fermi liquid ($p=\uparrow,\downarrow$), and for small
electron and hole concentrations ($n<2/3$ and $2-n<2/3$) the Fermi-liquid
component gives the main contribution (``overdoped regime'' of
high-$T_c$'s), whereas in other cases the non-Fermi liquid one
(``underdoped regime'').

\section{Acknowledgement}

This work was partially supported by the Fundamental Researches
Fund of the Ministry of Ukraine for Science and Education (Project
No.~02.07/266).

\parsep=0pt

\onecolumn

%\clearpage
%
%\section*{Figure captions}
%
%\begin{description}
%\item[Fig. 1:] Spectral weight function $\rho_{\sigma}(\omega)$
%    (\protect\ref{SWFt}): total and for each subspace, for the different
%    chemical potential values: (a) $\mu= U/2$, $n=1$; (b) $\mu=-1$, $n=0.07$;
%    (c) $\mu=0.01$, $n=0.72$; (d) $\mu=-0.01$, $n=0.66$ ($U=4$, $T=0.2$)
%\item[Fig. 2:] Statistical weights of the subspaces $w_p$ (\protect\ref{eq38})
%    as functions of the electron concentration ($U=4$, $T=0.2$)
%\item[Fig. 3:] Statistical weights of subspaces $w_p$ and ferromagnetic $m_{F}$
%    and antiferromagnetic $m_{AF}$ order parameters
%    vs electron concentration for $U=1.56$, $T=0.14$
%\item[Fig. 4:] Phase diagram $(T,U)$ at half-filling $n=1$ (AF~--
%    antiferromagnetic phase, PM~-- paramagnetic phase)
%\end{description}

\clearpage

%%Figure 1

%%Figure 2

%%Figure 3

%%Figure 4

\end{document}